\documentclass[iop,twocolappendix,numberedappendix]{emulateapj}

\usepackage{natbib}
\usepackage[loose, ugly]{units}
\usepackage{amsmath}
\usepackage[english]{babel}
\usepackage{blindtext}
\usepackage{hyperref}

\newcommand{\ceff}{\text{$c_{\rm eff}$}}
\newcommand{\hii}{\text{$\textrm{H}~\textsc{ii}$}}
\newcommand{\hi}{\text{$\textrm{H}~\textsc{i}$}}
\newcommand{\hmol}{\text{$\textrm{H}_2$}}
\newcommand{\stromgren}{Str\"{o}mgren}
\newcommand{\sbnn}{\texttt{starburst99}}
\newcommand{\msun}{\text{$M_\odot$}}

\shorttitle{Gravitationally Unstable Disk Galaxies II}
\shortauthors{N. J. Goldbaum, M. R. Krumholz, and J. C. Forbes}

\addto\extrasenglish{%

}

\begin{document}

\title{Mass Transport and Turbulence in Gravitationally Unstable Disk
  Galaxies. II:\\ The Effects of Star Formation Feedback}

\author{Nathan J. Goldbaum \altaffilmark{1}}
\affil{National Center for Supercomputing Applications, University of Illinois
  at Urbana-Champaign, 1205 W. Clark St., Urbana, IL 61801, USA}
\altaffiltext{1}{ngoldbau@illinois.edu}

\author{Mark R. Krumholz}
\affil{Research School of Astronomy \& Astrophysics,
  Australian National University, Canberra, ACT 2601, Australia and\\ Department
  of Astronomy \& Astrophysics, University of California, Santa Cruz, CA 95064,
  USA}

\author{John C. Forbes}
\affil{Department of Astronomy \& Astrophysics, University of California,
    Santa Cruz, CA 95064, USA}

\begin{abstract}
  Self-gravity and stellar feedback are capable of driving turbulence and
  transporting mass and angular momentum in disk galaxies, but the balance
  between them is not well understood. In the previous paper in this series, we
  showed that gravity alone can drive turbulence in galactic disks, regulate
  their Toomre $Q$ parameters to $\sim 1$, and transport mass inwards at a rate
  sufficient to fuel star formation in the centers of present-day galaxies. In
  this paper we extend our models to include the effects of star formation
  feedback. We show that feedback suppresses galaxies' star formation rates by a
  factor of $\sim 5$ and leads to the formation of a multi-phase atomic and
  molecular ISM.\@ Both the star formation rate and the phase balance produced
  in our simulations agree well with observations of nearby spirals. After our
  galaxies reach steady state, we find that the inclusion of feedback actually
  lowers the gas velocity dispersion slightly compared to the case of pure
  self-gravity, and also slightly reduces the rate of inward mass
  transport. Nevertheless, we find that, even with feedback included, our
  galactic disks self-regulate to $Q \sim 1$, and transport mass inwards at a
  rate sufficient to supply a substantial fraction of the inner disk star
  formation. We argue that gravitational instability is therefore likely to be
  the dominant source of turbulence and transport in galactic disks, and that it
  is responsible for fueling star formation in the inner parts of galactic disks
  over cosmological times.
\end{abstract}


\section{Introduction}

The star forming properties of isolated disk galaxies are driven by two primary
effects: gravitational instability and star formation feedback. Both can produce
supersonic turbulent motions in an initially laminar disk of gas. This
turbulence in turn can lead to non-axisymmetric stresses that mix the
interstellar medium (ISM) and transport mass inward and angular momentum
outward. We argued in \citet[hereafter Paper I]{goldbaum15a} that such mixing
and transport must be an essential component of any explanation for the
present-day properties of disk galaxies (also see similar arguments in, for
example, \citet{olivier91}, \citet{ferguson01}, \citet{krumholz10},
\citet{forbes12, forbes14}, and \citet{petit15}). In particular, most
present-day Milky Way-sized galaxies lack central holes in their gas
distributions \citep{bigiel12}, despite the fact that the gas consumption time
in their central regions is much shorter than a Hubble time \citep{bigiel08,
  leroy13}, and that most gas accretion either from a hot halo or from the
cosmic web is expected to arrive at large galactocentric radii, far from the
regions of active star formation \citep[e.g.,][]{dutton12,
  fraternali13}. Metallicity gradients in the gas dominated-outer regions of
nearby disks are also far too flat to be explained without redistribution of
metals from smaller galactocentric radii \citep{bresolin09, bresolin12, werk11,
  yang12, petit15}.

Because their effects are similar, it can be difficult to disentangle whether
star formation feedback or gravitational instability provides the dominant
explanation for any particular aspect of galaxy structure and evolution. In a
few cases the attribution is clear. For example, the gas velocity dispersions of
galaxies do not decline substantially outside $r_{25}$ \citep{tamburro09,
  ianjamasimanana12, ianjamasimanana15}, despite the almost complete absence of
star formation at such large galactocentric radii, a clear sign that star
formation feedback cannot be the key physical process there \citep{agertz09,
  agertz15, krumholz10, bournaud10, forbes12, forbes14}. Conversely, it has long
been established by simulations of disk galaxies that gravity alone is incapable
of destroying star forming clouds or producing enough turbulence to yield star
formation rates as low as those commonly-observed, strongly hinting that star
formation feedback is required \citep[e.g.,][]{tasker08, dobbs11b, renaud13,
  bonnell13}.

In many other cases, however, the effects are not so easy to separate, leading
to significant confusion. For example, one class of theoretical models assumes
that the velocity dispersion and/or \citet{toomre64} $Q$ parameter in galactic
disks is regulated by star formation feedback, and use this assumption to deduce
a star formation law \citep[e.g.,][]{thompson05, ostriker10, ostriker11,
  faucher-giguere13}. Another class of models assumes that $Q\sim 1$ is
maintained by gravitational instability, independent of the star formation law,
and instead derive the rate of mass transport through galaxies from this
assumption \citep[e.g.,][]{krumholz05, dekel09b, krumholz10, cacciato12,
  forbes12, forbes14, agertz15}. Clearly both pictures cannot be entirely
correct.

In Paper I, we presented a series of numerical experiments including
self-gravity but no stellar feedback, in order to isolate the role of
gravitational instability in determining the structure and evolution of disk
galaxies. We found that without feedback, star formation rates in our
simulations were roughly an order of magnitude too large compared to observed
star formation rates. However, we found that our model galaxies nonetheless
equilibrated to $Q \sim 1$, and to velocity dispersions of $\sim 10$ km
s$^{-1}$. The energy required to maintain this velocity dispersion and offset
the energy loss in radiative shocks came from accretion through the disk of the
galaxy. Our fiducial, Milky Way-like model produced a mass inflow rate of
$\sim 1$ \msun\ yr$^{-1}$ throughout the bulk of its disk, sufficient to fuel
all the star formation that is observed to occur in the inner disk of a Milky
Way-like galaxy. We therefore concluded that gravitational instability alone is
capable of fully explaining the observed velocity dispersions and marginal
gravitational instability of present-day galactic disks, and that it can explain
why star formation does not typically quench in their centers.

In this paper we present a new set of simulations using the same initial
conditions presented in Paper I, but now including a prescription for star
formation feedback. Our goal is to determine how the inclusion of feedback
modifies galactic structure and mass transport rates compared to the no-feedback
case. Below, we briefly discuss the initial conditions for our simulations
(\autoref{feed_ics}) and describe the theoretical basis and implementation for
our feedback model (\autoref{feedback_model}), which includes supernovae,
stellar winds, and photoionization-driven bubbles. This is followed by a
discussion of our simulation results, beginning with an overview of the
qualitative outcome of our simulations
(\autoref{feedback_qualitative_outcome}). Next, we discuss the impact of
feedback on the star formation rates and star formation histories
(\autoref{feedback_star_formation}) and ISM structure
(\autoref{feedback_ism_structure}) of our model galaxies. Finally, we
investigate the evolution of the gravitational instability in our model disks,
focusing on the gas velocity structure (\autoref{feedback_velocity_structure}),
Toomre $Q$ parameter (\autoref{feedback_gravitational_instability}), and rate of
radial mass transport (\autoref{feedback_mass_transport}).  We end by reviewing
our results and discussing them in context of galaxy formation and models for
galactic star formation rates (\autoref{discussion_conclusions}).

\section{Methods}

\subsection{Initial Conditions and Evolution}

\label{feed_ics}

To ease comparison with the models run without star formation feedback, in this
paper we will be discussing simulations initialized identically to the
simulations described in Paper I. This means that the initial portions of the
simulations are practically identical to the no feedback cases, up to the
formation of the first star particle --- any differences are due floating point
noise being amplified by dynamical chaos. Briefly, the model galaxies are
initialized using the \texttt{makegalaxy} code \citep{springel05}. The code
makes use of the analytic framework of \citet{mo98} to predict the properties of
a disk formed in a $\Lambda$CDM cosmology given a halo mass, disk mass, and halo
spin parameter.

The initial conditions include dark matter and stars, which are modeled as
N-body particles, and gas defined on an AMR mesh. The dark matter particles are
distributed according to a Hernquist profile, while the stars are distributed in
a thin exponential disk population and centrally concentrated bulge population.
Particle initial conditions are generated by randomly sampling from an analytic
distribution function, while the gas is initialized on the AMR mesh
following an analytic azimuthally symmetric exponential density profile.

The parameters of our model galaxies are chosen to loosely match the Milky Way,
with a halo mass of \unit[$\sim 10^{12}$]{\msun}, a stellar disk mass of
\unit[$\sim 10^{10}$]{\msun}, and a gas mass of \unit[$\sim
10^{9}$]{\msun}. We initialize three different galaxy models with identical
parameters besides the initial gas fraction. The low gas fraction (LGF) model
begins with a gas fraction of 10\% (relative to the mass of the stellar disk),
the fiducial model has a gas fraction of 20\%, and the high gas fraction (HGF)
model has an initial gas fraction of 40\%. We run the LGF and fiducial simulations
for 600 Myr. Due to numerical limitations, we have only evolved the HGF run
for 300 Myr.

The gas, stars, and dark matter are evolved using the Enzo code
\citep{enzo14}. The gas is evolved using second-order accurate PPM
hydrodynamics, and the gravitational potential is evaluated using a multigrid
method and particle dynamics are evolved using a kick-drift-kick scheme. The
maximum spatial resolution of the simulation is \unit[20]{pc}, and the
gravitational force resolution is two spatial elements, or \unit[40]{pc}. 

\subsection{Star Formation Feedback Model}

\label{feedback_model}

Here we describe a novel subgrid model for star formation feedback that includes
the effects of ionizing radiation from young stars, winds from evolved massive
stars, and the energy and momentum released by individual supernova
explosions. The parameters for the model are chosen to match a \sbnn\
\citep{leitherer99, vazquez05, leitherer14} model of a young stellar population
that fully samples the initial mass function (IMF). Additional details regarding
the model can be found in Forbes et al.\ (2016, submitted).

\subsubsection{Stochastic Stellar Population Synthesis}

The first step in our feedback model is the formation of a star particle. The
recipe for particle formation is the same as that used in Paper I, so we will
not repeat it here. All star particles in our simulations form with a uniform
initial mass of \unit[300]{\msun}. Within each of these particles we expect
there to be a few stars massive enough to produce supernova explosions. We model
this based on a scaled down version of a \sbnn\ model of a \unit[$10^6$]{\msun}
stellar population. For such a massive cluster, we should expect $\sim 10^4$
Type II supernovae (SNe) over the lifetime of the population. Scaling down to
our star particles, we expect to produce $\sim 3$ SNe per star particle. The
true number of SNe for each particle is chosen randomly at runtime, by drawing
from a Poisson distribution with an expectation value set by scaling the $10^6$
$M_\odot$ \sbnn\ model down to our 300 $M_\odot$ particle mass.

Once we have determined the number of SN explosions the star particle will
launch over the course of the simulation, we set the times at which each
individual SN will detonate by drawing from the supernova delay time
distribution predicted by \sbnn. There is a one-to-one mapping between supernova
delay time and progenitor initial mass, which we make use of to associate a
stellar mass to each SN progenitor. For each SN progenitor, we also record the
expected main sequence ionizing luminosity \citep{par03} (implicitly assuming
that massive stars that are below the minimum mass for a Type II SN explosion
contribute negligibly to the ionizing radiation budget) which we in turn use for
our \hii\ region feedback model. In practice, all of this data is regenerated at
each timestep using unique random number streams seeded by the unique particle
ID.\@ This adds extra CPU cost in each timestep while avoiding the memory and
communication costs of saving and synchronizing this data for all dynamically
created star particles.

Once the massive stars in each star particle have been identified, we loop over
all particles in the simulation, applying various forms of feedback to the
hydrodynamic quantities defined on the Enzo AMR mesh.

\subsubsection{\hii\ Region Feedback}

Massive stars emit copious amounts of ionizing radiation, capable of creating
bubbles of warm, ionized gas in the densest regions of the ISM.\@ The heating
provided by \hii\ regions can stabilize gas against gravitational collapse,
reducing the star formation rate immediately and directly.

We include \hii\ region feedback in our simulations using the following
algorithm. If a grid cell contains a dynamically created star particle
containing unexploded massive stars, we increase the gas temperature to emulate
photoionization heating. The amount of heating is determined by comparing the
volume of the \stromgren\ sphere, $V_s = (4/3) \pi R_s^3$, with the volume of
the cell in which the star particle resides, $V_c = \Delta x^3$, where
$\Delta x$ is the local cell spacing. Here,
\begin{equation}
R_s = \left( \frac{3 S}{4 \pi \alpha_B n^3} \right)^{1/3}
\end{equation}
is the \stromgren\ radius, determined by equilibrium between photoionization and
recombination at a given density, $S$ is the ionizing luminosity emitted by the
star particle, $n$ is the number density of gas in the host cell, and
$\alpha_B = 2.6 \times 10^{-13}\ \unit{cm^3/s}$ is the case-B recombination
coefficient, assuming a temperature of \unit[$10^4$]{K}. If $V_s \ge V_c$, the
cell is heated to a temperature of \unit[$10^4$]{K}. If the cell is already
hotter than that, no heating is applied. If $V_s < V_c$, we apply a volume
filling factor correction, heating the cell to $T = \unit[10^4]{K} \,(V_s/V_c)$.
If more than one particle contains SN progenitors in any given cell, they each
contribute separately, increasing the cell temperature up to a maximum of
\unit[$10^4$]{K}.

One major downside of this approach is that it is fully local to any given cell
in the simulation. In principle, \hii\ regions may grow to be larger than a
single cell, suppressing \hii\ region feedback and making it more difficult for
heated cells to dynamically affect the state of the simulation. In practice this
is not a big concern for the simulations we discuss here, since only the rarest
massive star clusters will produce \hii\ regions with diameters bigger than
\unit[20]{pc}. In anticipation of future simulations at higher resolution, we
plan to modify this feedback routine to apply \hii\ heating feedback in a
distributed fashion.

\subsubsection{Winds from Massive Stars}

In addition to heating by \hii\ regions, very massive stars eject a substantial
fraction of their envelopes, recycling gas back into the ISM.\@ We include this
effect in our feedback algorithm by altering the density and internal energy of
gas in cells that contain star particles associated with very young stellar
populations. We use the output of a \sbnn\ calculation to determine
the rate of wind energy and mass injection per unit mass and as a function of
age for a stellar population that fully samples the IMF.\@ We then add the
appropriate amount of mass and internal energy to each star particle's host cell
each time step, until the last SN explosion occurs.

The winds suppress star formation somewhat, since soon after a star particle
forms, the gas in the cell where the particle spawned will heat up. In addition,
this captures a significant fraction of the gas recycling from a young stellar
population. We do not include the effect of AGB winds, although that is not a
significant concern given that AGB stars only become significant after a few Gyr
and we evolve our simulations for a maximum of \unit[600]{Myr}. In practice, the
star formation suppression effect of winds is minor compared to SN explosions.

\subsubsection{Supernova Explosions}
\label{sn_feedback}
SN explosions are a commonly used source of feedback in simulations of galaxy
formation and evolution, including simulations on cosmological scales
\citep{cen92, springel03, scannapieco06} all the way down to parsec scales
\citep{joung06, kim11, kim15}. It has been known for many years \citep{katz92}
that simply depositing $E_{\rm SN} = \unit[10^{51}]{erg}$ per SN as thermal
energy in the location of the SN explosion does not produce strong feedback in
low to moderate resolution simulations. Instead, the energy, which must at
minimum be shared by all of the gas in a single computational element (typically
\unit[10 -- 1000]{pc} across) is quickly radiated away since the 
energy from a single SN is not enough to heat the amount of gas enclosed in such
an extended region above \unit[$\sim 10^6$]{K}. Since the cooling time is much
shorter at lower temperatures, the gas quickly radiates away the deposited thermal
energy before the supernova is able to do work on the surrounding
gas. At high ($\sim \unit[1]{pc}$) resolution this is less of a concern, but
simulating galaxy evolution at such high resolution is extremely expensive.

To circumvent this problem, low resolution simulations typically deposit the SN
energy continuously over many timesteps instead of using discrete
energy-injection events \citep{smith11}, possibly also using direct momentum
deposition to ensure gas is ejected out of halos \citep{oppenheimer08}.  The
combined heating of many SN explosions over millions of years is sufficient to
heat the gas in low resolution simulations, at the cost of resolving the
fine-grained structure of the SN feedback. In higher resolution simulations such
as ours, some have opted to directly resolve the SN blastwaves \citep{joung06,
  joung09}, temporarily turn off cooling in regions near the blastwave site to
avoid over-cooling \citep{stinson06}, directly depositing momentum into the
simulation \citep{kim11}, or using a hybrid of approaches \citep{kimm14,
  kimm15}.

Our approach most closely resembles the hybrid method of \citet{kimm14} and
\citet{kimm15}.  In our algorithm, each SN has an energy budget of
\unit[$10^{51}$]{erg}, but this energy is partitioned into two channels, which
we refer to as momentum feedback and energy feedback. 

The momentum feedback prescription is as follows. First, we identify particles
that will produce SNe in any given timestep. For each SN explosion, we identify
the grid cell where the exploding star particle resides and approximate that the
explosion happens in the center of the grid cell. This choice substantially
simplifies the implementation since we can exploit symmetry to naturally produce
a spherically symmetric explosion that conserves momentum by construction. In
the future, we plan to use the prescription described in \citet{simpson14},
which properly handles the problem of creating spherically symmetric
momentum-conserving blastwaves that are centered at an arbitrary location. In
practice our choice to recenter the explosion at the center of the host grid
cell leads to a small loss of resolution at scales below our grid spacing. We do
not expect a proper treatment of the sub-resolution position of the exploding
supernova to substantially alter the results discussed below.

Once the cell at the center of the supernova bubble has been identified, we
identify the 26 nearest neighbor cells and loop over each neighbor cell,
depositing momentum
\begin{equation}
  \Delta p_{\rm SN} = 1.2 \times 10^4\ \unit{\msun}\ 
    \unit{km/s}\ \hat{r},
\end{equation}
where $\hat{r}$ is the unit vector connecting the center of the supernova host
cell to the neighbor cell under consideration. The normalization for
$\Delta p_{\rm SN}$ used here was chosen based on detailed simulations of SN
explosions in a turbulent ISM where the net momentum injection into the ISM can
be directly measured \citep[c.f.][]{cioffi88, kim11, kim15, martizzi15}. The
total momentum deposited is $3 \times 10^5\ \unit{\msun} \unit{km/s}$, shared
equally between each of the 26 neighbor zones. If the SN explodes at the edge of
a grid patch, the portion of the explosion that would happen in cells living on
a different grid is not included, effectively ``cutting off'' the supernova
explosion and abandoning spherical symmetry. This happens relatively rarely, but
is a deficiency of our feedback algorithm, albeit one that is commonly used in
distributed feedback calculations in the Enzo code \citep[although in the latter
the feedback region is shifted rather than cut off]{kim11b, simpson14}. To avoid
producing spuriously fast-moving gas when a supernova explodes in the
neighborhood of a relatively low-density cell, we limit the maximum change in
velocity for the gas in any given cell to \unit[1000]{km/s}.

Since particles that spawn supernova explosions must always live on the maximum
refinement level, the mass of cells in the neighborhood of stars that go
supernova is independent of the local gas temperature. This means that heating
from \hii\ region feedback and stellar wind feedback does not substantially
alter the dynamics of the blastwaves, since they are driven by direct momentum
injection. Instead, the primary effect of the stellar wind and \hii\ region is
to provide thermal support in the densest regions of the ISM and produce a more
realistic ISM structure (see \autoref{feedback_ism_structure} for a more
detailed discussion of the ISM structure in our simulations).

The energy feedback portion of our algorithm occurs once we have computed the
momentum to be added. We record the change in kinetic energy with respect to the
simulation's rest frame for each cell that we deposit momentum into. The total
net increase in kinetic energy in the cells surrounding the supernova host cell
is then deducted from the available budget of \unit[$10^{51}$]{erg} and the
balance of the energy is then deposited in the SN host cell as thermal
energy. We also adjust the mass and metallicity in the host cell to account for
mass injection due to the supernova ejecta. In practice, at \unit[20]{pc}
resolution, most of the thermal energy is used to launch the momentum feedback,
so the amount available for heating the host cell is much less
\unit[$10^{51}$]{erg}. We discuss how this affects our results in
\autoref{feedback_ism_structure}, when we note how the hot phase of the ISM is
suppressed compared to observed hot phases in nearby star forming galaxies. 

In a small fraction, $\sim 1\%$, of supernova explosions, the kinetic energy
injected by our momentum feedback prescription exceeds \unit[10$^{51}$]{erg}. In
these cases, we inject no energy into the SN host cell, but do not go back to
redo the momentum injection to ensure detailed energy balance. This happens when
a supernova explosion goes off in the neighborhood of one or more low-density
cells, so we do not expect this effect to impact our conclusions regarding the
destruction of dense star-forming clouds.

Before moving on, we pause to note that one advantage of our feedback
prescription is that, at least in its treatment of the cold ISM, it should not be
tremendously sensitive to resolution. This is because the effects of the gas on
the cold ISM come mainly from the momentum feedback, and the total amount
of momentum we are injecting is resolution-independent. Since momentum
cannot be radiated away, and nothing behaves non-linearly with density (as
does, for example, the rate of radiative cooling), the volume into which the
momentum is injected does not affect the results as strongly as it would for
an energy-driven feedback prescription. This statement does not apply to
the energy feedback portion of our prescription, however, a point that will
become relevant in \autoref{galactic_winds}.

\section{Results}

Here we describe the results of our simulations. First, in
\autoref{feedback_qualitative_outcome}, we focus on the qualitative evolution of
our simulated disks, focusing on their morphology. Next, in
\autoref{feedback_star_formation} we describe the impact of our feedback
algorithm on the star formation rates and star formation histories. In
\autoref{feedback_ism_structure} we describe the structure of the ISM in our
simulated galaxies, showing how feedback moderates the amount of gas available
for star formation by dispersing dense gas concentrations. Finally, in
\autoref{gravitational_instability} we discuss the development of gravitational
instability via measurements of the turbulent velocity structure, Toomre Q
parameter, and the mass transport rate.

Throughout this section, all the derived quantities we measure are computed
exactly as in Paper I, and we refer readers to the Appendix of that paper for a
full description of our analysis pipeline. The source code for the pipeline and
supplementary analysis scripts are available on
Bitbucket\footnote{\url{https://bitbucket.org/ngoldbaum/galaxy_analysis}}. The
raw simulation data, initial conditions, and ancillary postprocessed data are
also available for download\footnote{\url{http://dx.doi.org/10.13012/J85Q4T1T}}.

\subsection{Qualitative Outcome}

\label{feedback_qualitative_outcome}

\begin{figure*}
\plotone{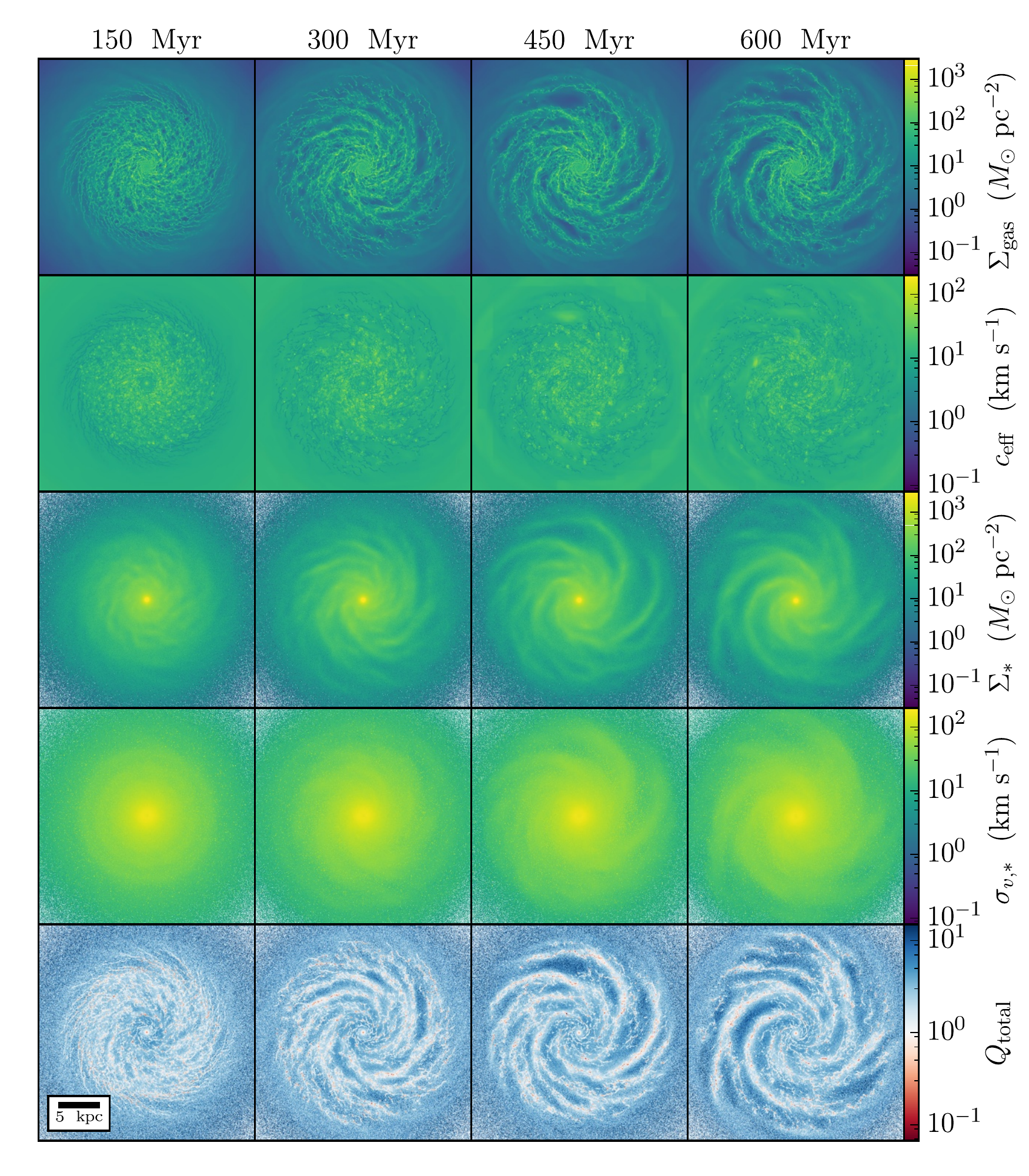}
\caption{The surface density and vertically-averaged effective sound speed for
  both the gas and stars, as well as the combined Toomre $Q$ parameter after
  \unit[150, 300, 450, and 600]{Myr} has elapsed in the fiducial simulation.}
\label{visual_evolve_summary_feedback}
\end{figure*}

We present snapshots of the fiducial simulation at four times in
\autoref{visual_evolve_summary_feedback}. The Figure displays stellar surface
density and effective sound speed for both the gas and stars, as well as
$Q_{\rm total}$, which includes contributions from both gas and stars. The
outcome of all three simulations are similar to the evolution depicted in
\autoref{visual_evolve_summary_feedback}, so we will describe the features
shared between the three simulations here and explain how the results vary as a
function of gas fraction below.

Early in the simulation, the morphology is dominated by circular, expanding
rings of gas and stars, which we interpret as a manifestation of unphysical
transient behavior as the disk settles out of its simplified initial conditions. Later,
the rings dissipate, giving way to prominent spiral arms. The arms form
spontaneously, and are clearly visible in both the gas and stellar surface
density maps.

The gas in the inner disk is concentrated in thin filaments that are
continuously sheared, leading to a tightly wound spiral pattern. At intermediate
radii, the filaments break up into individual isolated clouds. This morphology
develops as a smooth transition along the prominent spiral arms at the edge of
the star forming disk. Gas in the outer disk is smoother, only collapsing into
more or less isolated clouds towards the end of the simulation. Individual SN
explosions may also temporarily evacuate the area around the explosion
site. This is visible in the gas density, in the upper right quadrant of the
disk in the snapshot at \unit[$T=300$]{Myr} in
\autoref{visual_evolve_summary_feedback}, where several blastwaves have
evacuated regions embedded in a spiral arm.

\begin{figure*}
\plotone{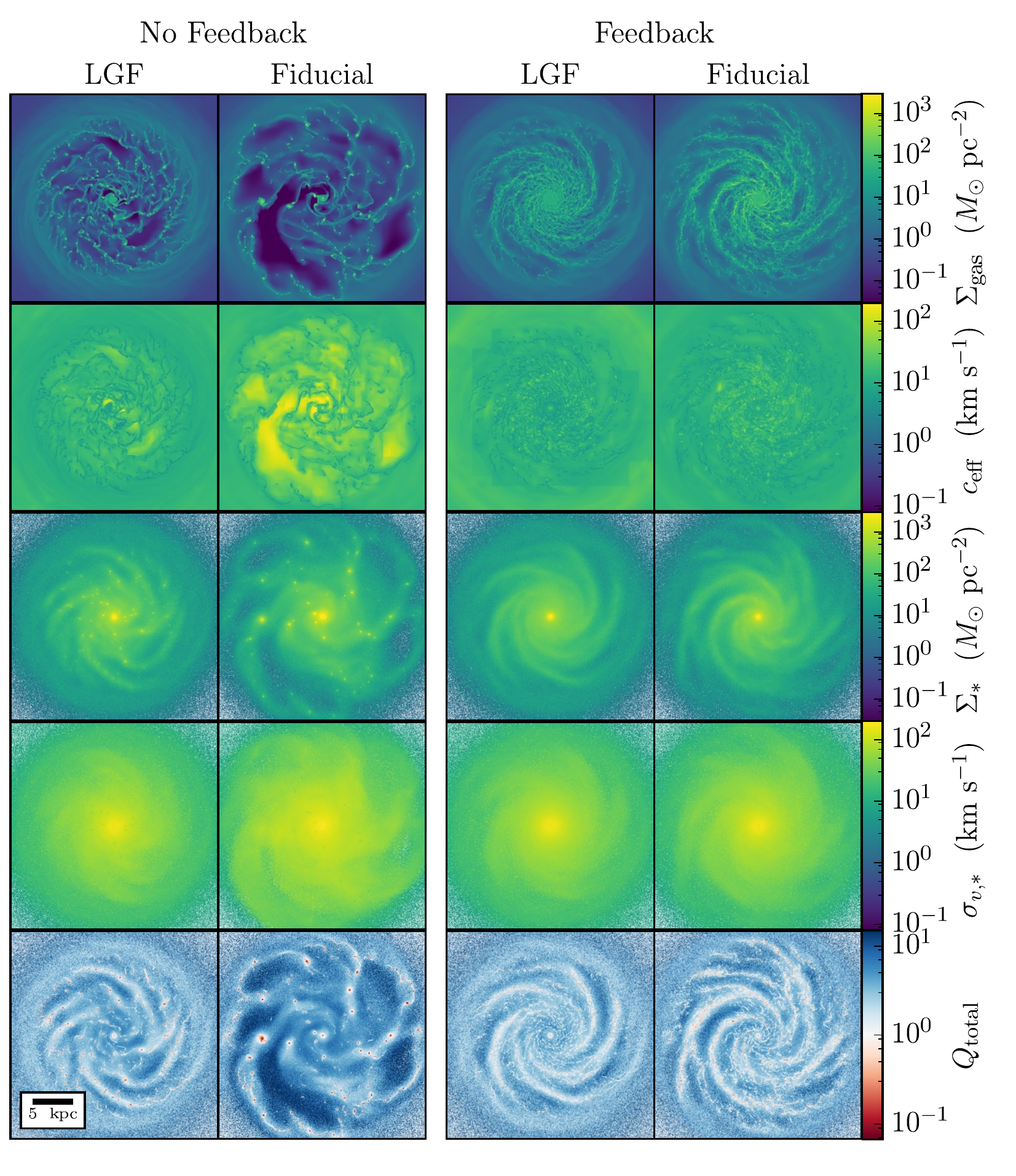}
\caption{Same as \autoref{visual_evolve_summary_feedback}, but showing a fixed
  simulation time ($T = \unit[600]{Myr}$) for four simulations. We show both the
  low gas fraction (first and third columns) and fiducial (second and fourth
  column) cases with (left two columns) and without feedback (right two
  columns). The high gas fraction simulations are not included since the run
  including star formation feedback was only evolved for \unit[300]{Myr}. The
  simulations with no feedback are described in Paper I.}
\label{visual_summary_sims_feedback}
\end{figure*}

The gas effective sound speed is roughly constant with radius, in an azimuthally
averaged sense, although with significant variation at any given
radius. Individual SN explosions produce a substantial amount of hot gas
in the inner disk, with \ceff\ as high as \unit[$\sim$ 50--100]{km/s} in
interarm regions. Dense gas collects in filaments, where \ceff\ is suppressed
compared to the interarm regions, with \ceff\ \unit[$\sim\ $ 1--10]{km/s}. We
will show in \autoref{feedback_velocity_structure} that the effective sound
speed in the dense gas is dominated by the turbulent velocity dispersion. In the
outskirts of the galaxy, SN-heated regions become rarer. Here the
interarm gas is still heated to \unit[10--20]{km/s}, and appears similar to the
interarm gas seen in the simulations without feedback.

The stellar component is substantially smoother, albeit morphologically similar
to the gas on \unit{kpc} scales. Once the disk has settled, the stars show a
prominent $m=5$ spiral pattern. The stellar velocity dispersion does not vary
much over the course of the simulation, showing neither appreciable heating nor
dissipation. For this reason, the distribution of $Q_{\rm total}$ is primarily
dictated by the behavior of the gas. Early in the simulation, as the disk comes
into equilibrium, $Q_{\rm total} \simeq$\ 1--2, with $Q_{\rm total} \lesssim 1$
in the highest gas density regions and $Q_{\rm total} \simeq 1$ in the interarm
regions.  Later in the simulation, $Q_{\rm total}$ increases in the interarm
regions, reaching as high as $\sim 5$ in inter-arm regions, while remaining
$\sim 1$ within the spiral arms.

We can see the profound effect of feedback on the structure of our simulated
galaxies by comparing directly with the runs with no feedback, as shown in
\autoref{visual_summary_sims_feedback}. At late times in the runs with no
feedback the morphology is dominated by long-lived bound clumps, while in the
runs with feedback the gas is distributed relatively smoothly, although it is
modulated by the spiral pattern.  Since gravitationally bound clouds are
efficiently destroyed by supernova feedback, the stellar distribution is also
substantially more smooth, showing no clear concentrations of stars besides the
bulge population present in the initial conditions. The gravitational stability
parameter $Q_{\rm total}$ is also more smoothly varying in the runs with feedback

When comparing the low gas fraction and fiducial simulations with feedback, we
see the disks look very similar. The star forming region is smaller in the low
gas fraction case since the outskirts of the disk have not had time to collapse,
but the inner disks are practically indistinguishable. We do see some
differences in the effective sound speed, which is somewhat higher in the inner
portion of the fiducial simulation. This difference between the fiducial and LGF
simulation is driven by the substantially higher star formation rate in the
fiducial simulation, which produces more gas heated by both photoionization and
supernova thermal feedback.

None of our simulations produce significant galactic winds. This may be due to
the resolution of our models, but might also be due to the masses of our
simulated disks. We discuss this point in more detail in
\autoref{galactic_winds}.

\subsection{Star Formation}

\label{feedback_star_formation}

\begin{figure}
\plotone{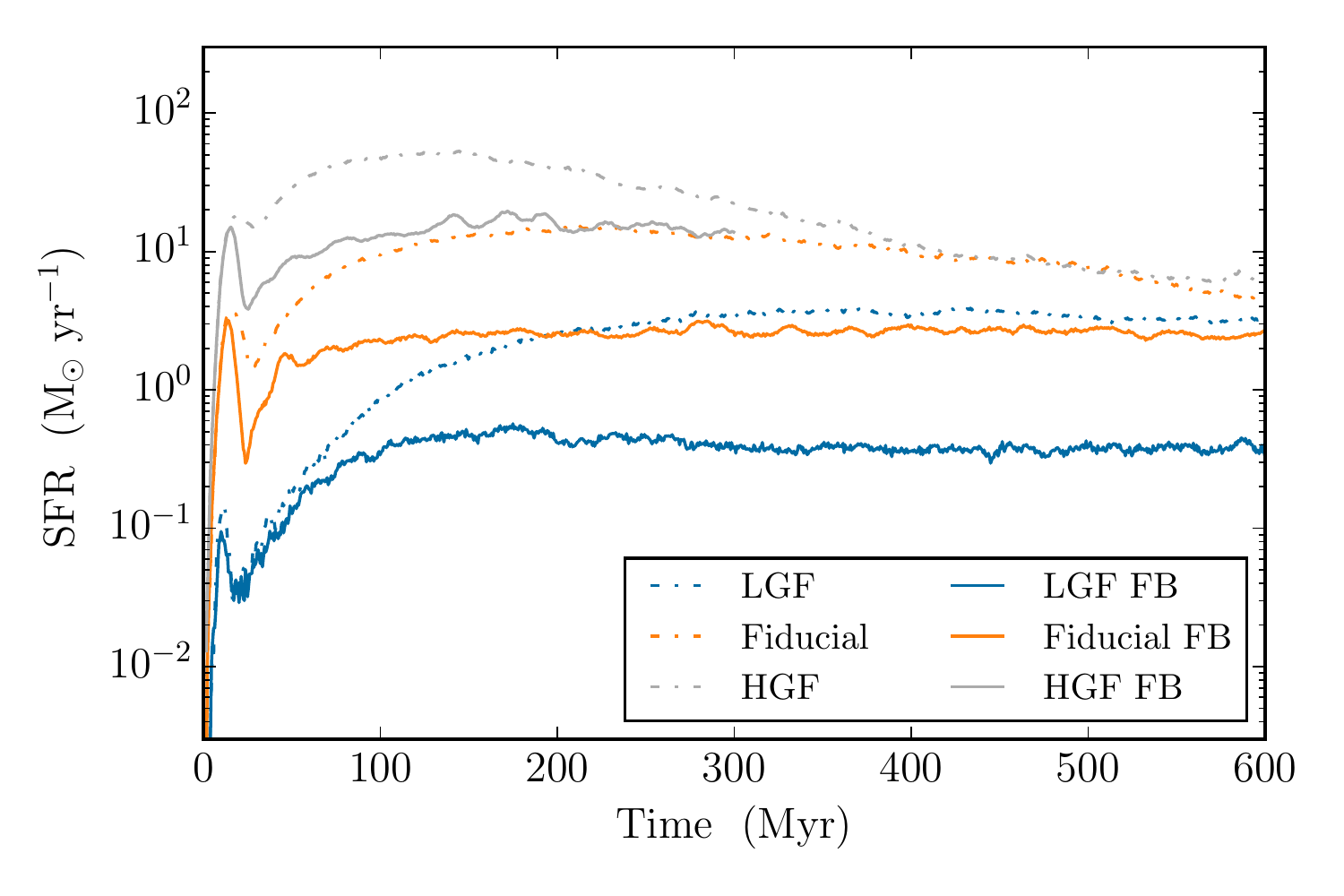}
\caption{The star formation rate history for our simulated galaxy models. We
  show the low gas fraction (blue), fiducial (orange), and high gas fraction
  (gray) cases, both with (solid lines) and without (dashed lines) star
  formation feedback.}
\label{sfr_figure_feedback}
\end{figure}

Star formation feedback has a profound effect on the star formation rates in our
model galaxies. In \autoref{sfr_figure_feedback}, we show the star formation
history of our galaxy simulations as measured by binning the dynamically created
star particles present at the end of each simulation by creation time, weighting
by the initial particle mass. We show models run with and without feedback for
all three choices of initial gas fraction.

Initially, the models are identical: the star formation rate shows a peak
followed by a decline. This pattern is driven by the dynamical nature of the
initial collapse of the disk, which is identical in the cases with and without
feedback up to the formation of the first star particle. However, in both the
fiducial and high gas fraction cases, the initial peak is somewhat depressed and
the following trough in the star formation history is deeper.  In the low gas
fraction case the star formation histories agree for a somewhat longer period,
diverging only as the star formation rate begins to increase again.

\begin{figure}
\plotone{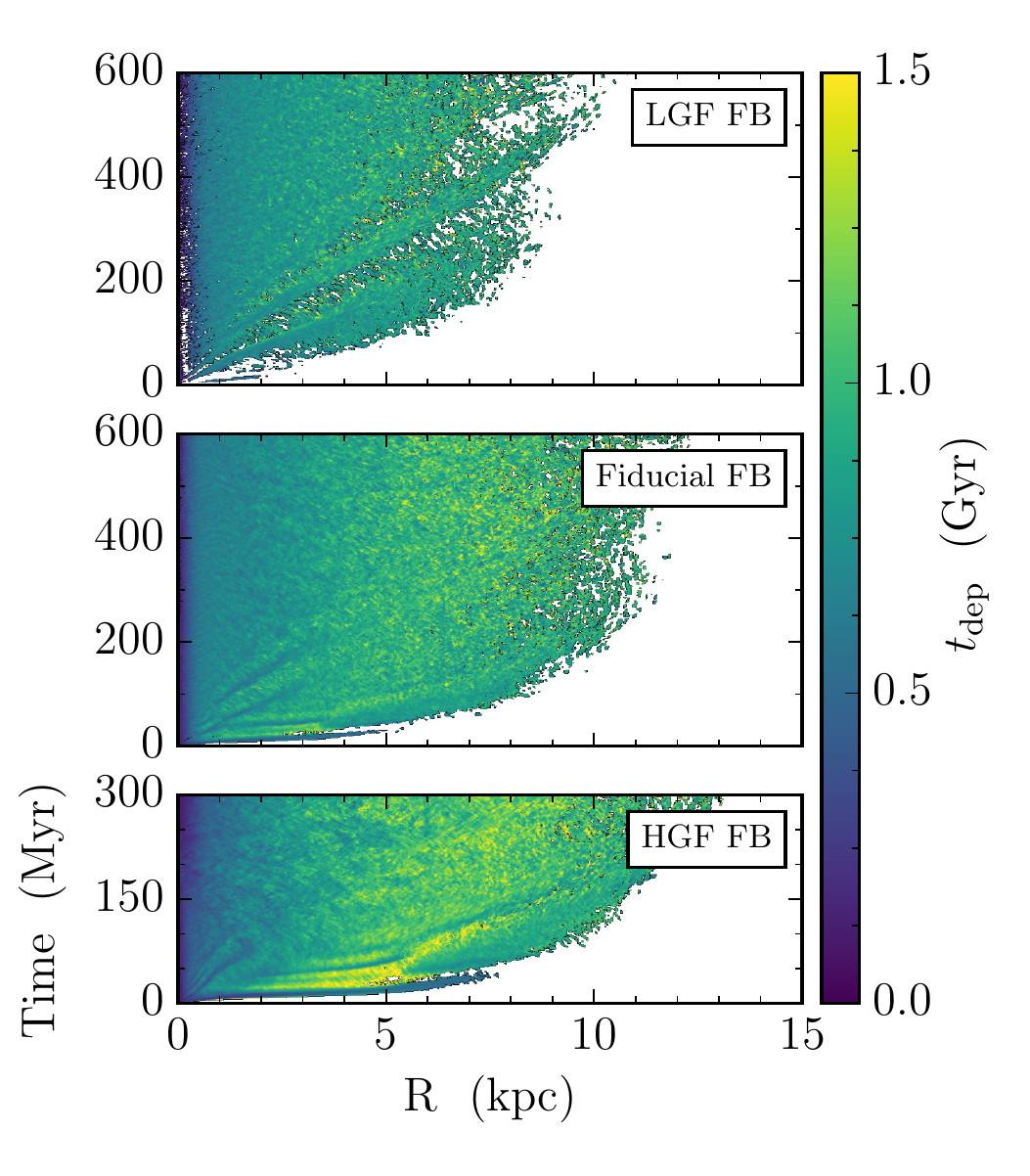}
\caption{The azimuthally averaged gas depletion time ($t_{\rm dep} = \Sigma_{\rm
    gas}/\Sigma_{SFR}$) as a function of radius and time.}\
\label{radius_time_star_formation_depletion_time_feedback}
\end{figure}

Eventually, for the low gas fraction and fiducial cases, the star formation
rates in the simulations with feedback converge to a quasi-equilibrium
value. The high gas fraction case also appears to be converging to an
equilibrium value, although it has not run long enough to fully converge. For
the low gas fraction, fiducial, and high gas fraction cases, the star formation
rates converge to $\unit[\sim 0.3]{\msun\ {\rm yr}^{-1}}$,
$ \unit[\sim 2]{\msun\ {\rm yr}^{-1}}$, and
$\unit[\sim 10\text{--}20]{\msun\ {\rm yr}^{-1}}$, respectively. For the runs
with feedback, we do not see the long-term decrease in the star formation rate
seen in the runs with no feedback. Those decreases are primarily driven by gas
depletion, and since the star formation rate is substantially depressed in the
simulations with feedback, there is not sufficient time over the course of our
simulations to substantially deplete the gas.

The equilibrium star formation rate scales super-linearly in the gas
fraction. The superlinearity has two causes that we can identify. First,
as we discuss further in \autoref{feedback_ism_structure}, the runs
with higher gas fractions have higher fractions of their ISM in cooler
phases that lack thermal support and are liable to undergo star formation. This
is primarily a density effect: runs with higher gas fractions have higher
midplane densities, and this shifts the balance between radiative
cooling ($\propto n^2$) and heating ($\propto n$) more in favor of 
cooling. Second, a higher gas fraction also raises the midplane pressure,
and this in turn raises the density of the cold, star-forming gas. The
increase in density in turn lowers the dynamical time in this gas, allowing
it to form stars faster.

In comparing with observations, it is also helpful to consider the star
formation rate point by point, rather than summed over the entire galaxy. To
characterize the star formation behavior on such smaller scales, it is helpful
to consider the depletion time \begin{equation} t_{\rm dep} = \frac{\Sigma_{\rm
      gas}}{\Sigma_{\rm sfr}},
\end{equation}
the time it would take to consume all of the gas at a given position in a
galaxy. Resolved observations of nearby star forming galaxies indicate typical
depletion times in the molecular phase of $\unit[\sim 2]{Gyr}$, with
approximately a third of a decade of scatter \citep{bigiel08, bigiel11,
leroy13}.

We show the azimuthally-averaged depletion time as a function of 
radius and time in our simulations 
in \autoref{radius_time_star_formation_depletion_time_feedback}.
We see that, in comparison to observations, our model galaxies have somewhat lower
depletion times, with $t_{\rm dep} = \unit[0.5\text{--}1.5]{Gyr}$ throughout
most of the star forming disk over the full course of all three simulations. In
the inner galaxy, $t_{\rm dep}$ is lower, $\sim \unit[200]{Myr}$. There is some
observational evidence for such a decline in depletion time in the very centers
of galaxies \citep{leroy13}, but by less than occurs in our simulations. We also
note that the time evolution of the azimuthally averaged depletion time shows
relatively little structure, particularly in the low gas fraction case. There is
substantially more variation at a fixed time in our simulations, but performing an
azimuthal average washes out most of the extreme variations due to the formation
of star-forming clouds and evacuated inter-arm regions. In the fiducial case we
also see that the depletion time is suppressed somewhat in spiral arms,
particularly at early times. There is some observational evidence that molecular
gas in regions of high shear have longer depletion times \citep{meidt13a}, and
our simulations appear qualitatively consistent with this finding. However, we
note that mechanism proposed by \citet{meidt13a} for changing the depletion time
--- variations in the cloud surface pressure between low- and high-shear regions
leading to variations in the force balance within molecular clouds --- appears
unlikely to explain the results in our simulations, since our resolution is
insufficient to capture the internal structure of molecular clouds.

\begin{figure*}
\plotone{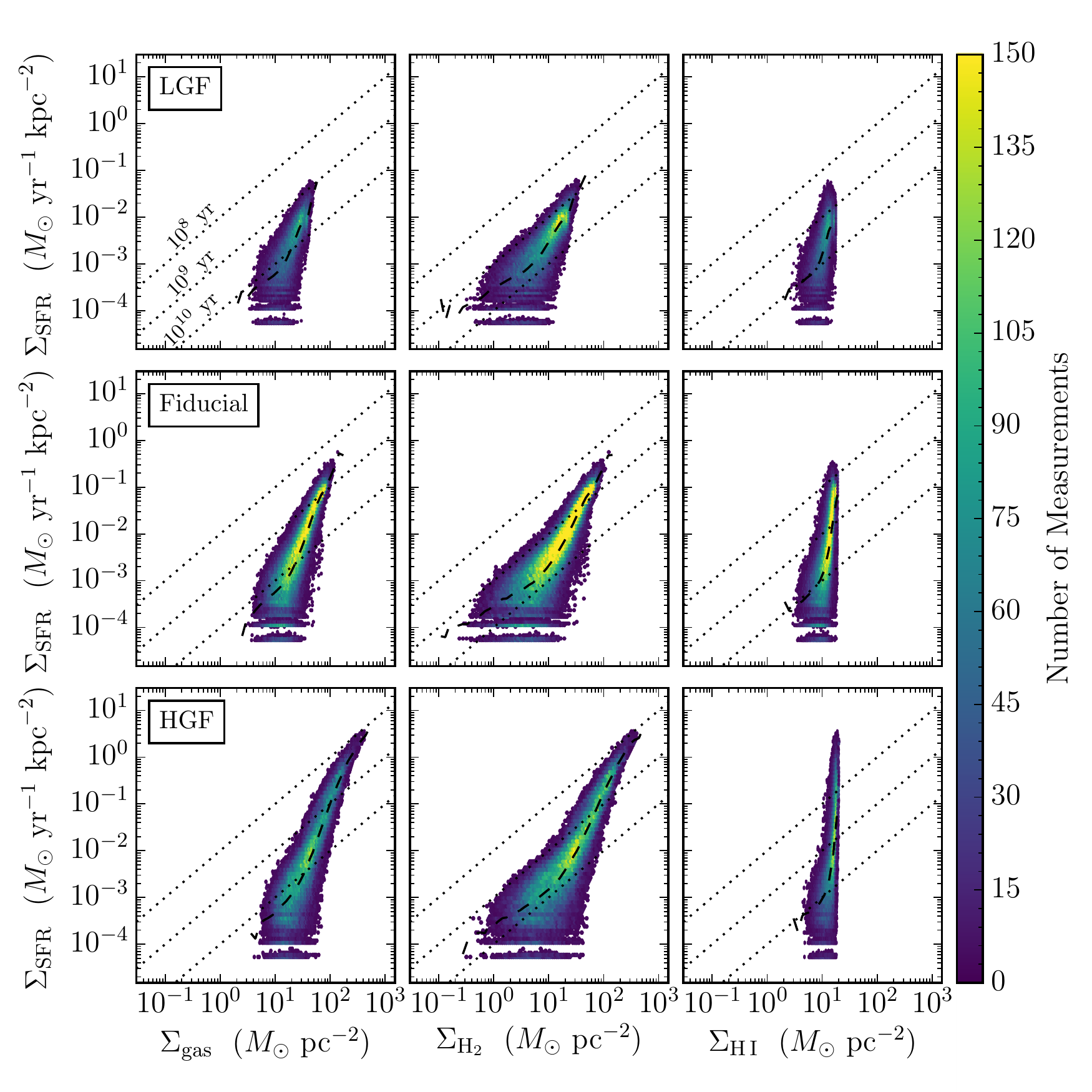}
\caption{The star formation rate surface density as a function of gas (left
  column) \hmol\ (middle column) and \hi\ (right column) surface density.  We
  show data from the low gas fraction (top row), fiducial (middle row) and, high
  gas fraction (bottom row) models. To guide the eye, we also mark lines of
  constant depletion time ($10^8$, $10^9$, and $10^{10}$ years) as dotted
  lines and the median of the measured relations as dashed lines.}
\label{ks_plot}
\end{figure*}

To guide our intuition, we can also directly compare to observations of resolved
star formation in nearby galaxies. This is shown in \autoref{ks_plot}, which can
be directly compared to Figure 4 of \citet{bigiel08}, Figure 1 of
\citet{leroy13}, or Figures 2 and 3 of \citet{krumholz14}.  To generate this
plot, we first generate maps of the \hmol\ and \hi\ surface densities by making
use of the analytic approximations of \citet{kmt08}, \citet{kmt09}, and
\citet{mckee10}. These approximations are based on detailed calculations of
self-shielding in idealized cloud complexes and can predict the molecular gas
fraction $f_{\hmol} = \Sigma_{\hmol}/\Sigma_{\rm gas}$ as a function of
$\Sigma_{\rm gas}$ and metallicity. In practice, we use the formula given in
Equation 93 of \citet{mckee10},
\begin{equation}
f_{\hmol} = 1 - \frac{0.75 s}{1 + 0.25s}
\end{equation}
where
\begin{equation}
\label{h2_equation}
s = \frac{\ln\left(1 + 0.6\chi + 0.01\chi^2\right)}{0.6\tau},
\end{equation}
$\chi = 0.77(1 + 3.1Z^{0.365})$,
$\tau = 0.066 Z \Sigma_{\rm gas}/(\msun\ \unit{pc^{-2}})$, and $Z$ is the
metallicity relative to the solar value. Our simulations are initialized at
solar metallicity and do not show significant metallicity evolution so we assume
$Z=1$ when evaluating \autoref{h2_equation}. We generate maps of
$\Sigma_{\rm SFR}$ by projecting the expected star formation rate along the
$z$-axis, given our star formation law.

We infer $\Sigma_{\hmol}$ via
\begin{equation}
\Sigma_{\hmol} = f_{\hmol}\Sigma_{\rm gas}
\end{equation}
and $\Sigma_{\rm HI}$ via,
\begin{equation}
\Sigma_{\rm HI} = \Sigma_{\rm gas} - \Sigma_{\hmol}.
\end{equation}
The latter expression ignores the contributions of ionized gas, which we will
show is a good approximation in \autoref{feedback_ism_structure}, and assumes
the mass fraction of metals is negligible.

To directly compare with observations, which are typically done with
$\sim \unit{kpc}$ scale resolution elements (although see \citealt{sch10} and
\citealt{ono10}), we degrade the resolution of our surface density maps to
\unit[740]{pc}, matching the resolution of the THINGS survey \citep{leroy08,
  bigiel08}. Next, we take each pixel at each simulation time in our degraded
resolution surface density maps to be an independent measurement of the surface
density of atomic hydrogen, molecular hydrogen, and star formation. We combine
the set of all measurements we infer from all of our simulation snapshots, and
create a 2D histogram, which we plot in \autoref{ks_plot}.

In all three cases, the classical Kennicutt-Schmidt law \citep{kennicutt98,
  kennicutt12}, where the total gas surface density is on the x-axis, shows a
super-linear scaling, as observed in nearby star forming galaxies. The molecular
gas Kennicutt-Schmidt law is shallower, showing a roughly linear scaling at the
low surface density end, moving to a superlinear scaling at the high surface
density end.  Lastly, as observed in nearby galaxies \citep{bigiel08}, the \hi\
surface density shows no correlation with the surface density of star formation.

In both the fiducial and high gas fraction simulations, we see a transition to a
low depletion time mode of star formation at the high gas surface density side
of the phase space. This may correspond to a ``starburst'' mode of star
formation \citep[cf.][]{daddi10}, or may simply indicate that our feedback
prescription is inefficient at destroying high surface density clouds, allowing
star formation to proceed there for longer than in lower surface density
conditions.

\subsection{ISM Structure}

\label{feedback_ism_structure}

As we showed in \autoref{feedback_qualitative_outcome}, particularly in
\autoref{visual_summary_sims_feedback}, our feedback model has a dramatic effect
on the equilibrium structure of the ISM in our model galaxies.  Rather than
ending up in a state where the bulk of the gas collects into massive
gravitationally bound clouds, the gas is instead more smoothly distributed
throughout the disk, only collecting in large-scale spiral arm patterns.

\begin{figure}
\epsscale{1.1}
\plotone{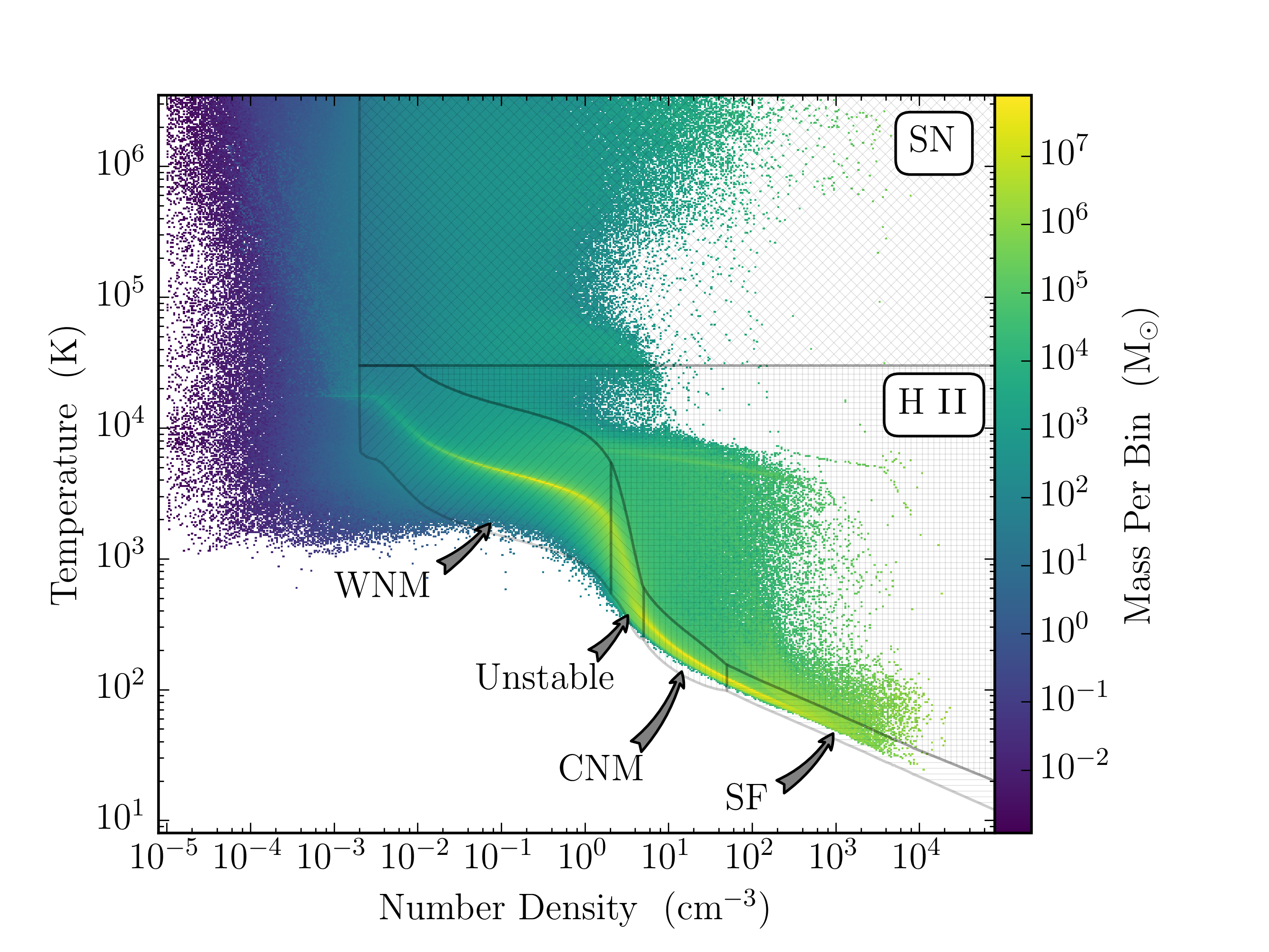}
\epsscale{1.0}
\caption{A density-temperature phase diagram for the gas in the fiducial
  simulation at T = \unit[600]{Myr}. We indicate the regions of phase space we
  use to define various components of the ISM.}
\label{phase_diagram_labeled}
\end{figure}

We can make this statement quantitative by segmenting the gas in our simulated
galaxies according to ISM phase. In \autoref{phase_diagram_labeled}, we show an
example temperature-density phase diagram, with various ISM phases marked as
cross-hatched regions. The regions we identify can primarily be separated into
two components: equilibrium and non-equilibrium phases. The former, which
include the warm neutral medium (WNM), thermally unstable phase (Unstable), cold
neutral medium (CNM), and star forming gas (SF), all fall along the equilibrium
cooling curve, i.e., the region of phase space where, at a fixed density,
cooling (primarily due to metal line emission) and heating (primarily due to
thermalization of photoelectrons ejected off of dust grains) exactly balance
\citep{fgh69, wolfire03}. In practice, these equilibrium regions have a finite
width due to small-scale density and temperature fluctuations, so we also
include a narrow vertical range above and below the equilibrium curve for each
equilibrium component. Non-equilibrium phases, including gas heated by \hii\
region feedback and supernova thermal feedback, fall above the equilibrium
cooling curve. Note that while we do not explicitly include a prescription for
the formation of molecular hydrogen, our star formation threshold corresponds
approximately to the transition to the molecular phase, and so our star forming
phase can be thought of as corresponding to the molecular gas phase in a real
galaxy.

\begin{figure}
\epsscale{1.1}
\plotone{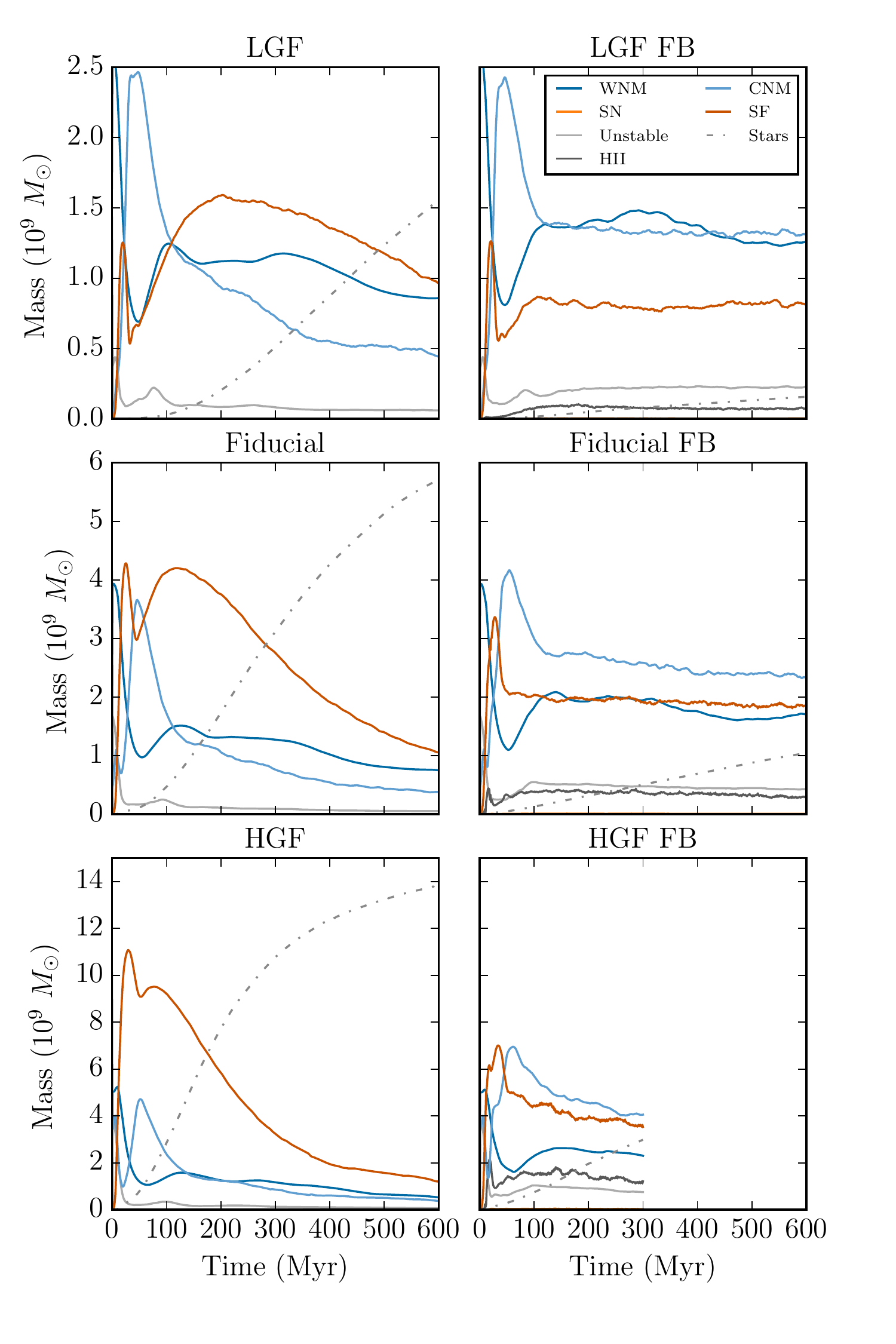}
\epsscale{1.0}
\caption{The time evolution of the masses of the ISM components depicted in
  \autoref{phase_diagram_labeled} (solid lines) along with the mass of
  dynamically formed stars (dot-dashed line). The left column shows simulations
  without feedback, while the right column shows those with feedback.}
\label{ism_mass_evolve}
\end{figure}

We can quantitatively compare the impact of star formation feedback on the
structure of the ISM by finding the gas mass in each ISM component as a function
of time for each of our simulated galaxies.  The results of this comparison are
shown in \autoref{ism_mass_evolve}.  We show simulations without (left column)
and with (right column) feedback, for each choice of initial gas fraction.

In all three simulations without feedback, the bulk of the gas mass is locked up
in dense star forming gas. Over the course of the simulation, this gas is
converted into star particles, until eventually the gas supply is
exhausted. While there is still substantial gas left at the end of the
simulation in the low gas fraction run, the bulk of the gas in fiducial and high
gas fraction is converted into stars over the course of the simulation.

The story is markedly different in the simulations with feedback.  Rather than
being locked up in star forming gas, the bulk of the ISM is in the WNM or CNM.\@
In the low gas fraction case, the gas is approximately evenly split between WNM
and CNM, with a smaller fraction ending up as star forming gas. In the fiducial
and high gas fraction cases, the bulk of the gas ends up in the CNM and
star forming gas is a substantially larger fraction of the ISM mass compared to
the LGF run. In both the fiducial and low gas fraction cases, the ISM develops
an equilibrium configuration, where the mass of each component is approximately
constant over timescales of several hundred \unit{Myr}. While the mass of gas
heated in the \hii\ heated phase is non-negligible in the runs with feedback,
this gas is small fraction of the overall mass of the ISM.\@ In all cases the
mass of SN heated gas is negligible.

\subsection{Gravitational Instability}

\label{gravitational_instability}

\begin{figure}
\plotone{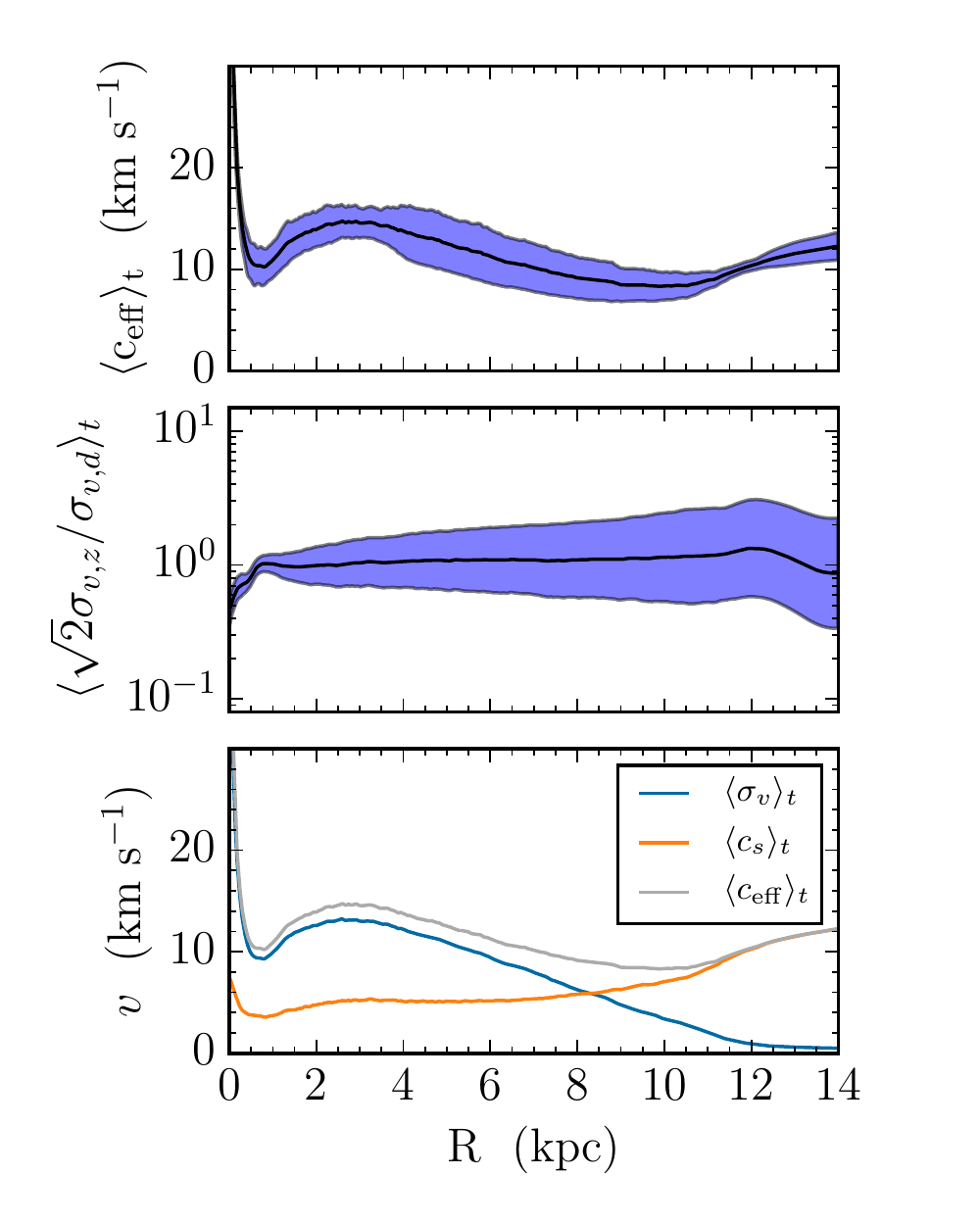}
\caption{The time-averaged gas effective sound speed (top panel), time-averaged
  velocity dispersion anisotropy (middle panel), and the contribution to the
  effective sound speed by the velocity dispersion and thermal sound speed
  (bottom panel) for the fiducial simulation. The blue shaded regions in the top
  two panels indicate 1-$\sigma$ variance of the plotted quantity as a function
  radius.}
\label{velocity_summary_plot_feedback}
\end{figure}

Here we focus on the gravitational instability that develops in our simulated
galaxies.  In \autoref{feedback_velocity_structure}, we focus on the velocity
structure in the gas. This is followed in
\autoref{feedback_gravitational_instability} by a discussion of the evolution of
the Toomre $Q$ parameter in our simulated disks, showing how an equilibrium
value of $Q_{\rm total}$ naturally develops. Lastly, in
\autoref{feedback_mass_transport}, we measure the mean radial mass transport
rate, and compare it with the star formation rate.

\subsubsection{Velocity Structure}

\label{feedback_velocity_structure}

As we showed above, the gas in our simulated galaxies undergoes a cycle of
collapse into gravitationally bound clouds, rarefaction due to supernova
feedback, followed by re-collapse into gravitationally bound clouds. Both
supernova explosions and local departures from a purely axisymmetric
gravitational potential generate substantial turbulent velocity dispersions.  In
addition, supernova explosions, winds from massive stars, and \hii\ regions can
heat the gas, providing support for the gaseous disk in the form of thermal
pressure.

We can see the typical velocity structure in the gaseous disk by inspecting
\autoref{velocity_summary_plot_feedback}, where we plot the time average of the
gas effective sound speed, sound speed, turbulent velocity dispersion, and the
anisotropy in the turbulent velocity dispersion as a function of galactocentric
radius.  At all radii, $\ceff \gtrsim \unit[8]{km/s}$, reaching as high as
\unit[15]{km/s} at $R = \unit[3]{kpc}$.  In the inner, star forming portion of
the disk, the effective sound speed is mostly due to turbulent motions, while
the sound speed dominates in the outer disk. The mass-weighted sound speed is
lower in the inner disk because a typical parcel of gas in the star-forming
inner disk will have a higher density, and thus be able to cool more effectively
than gas in the outer disk. Compared to the simulations without feedback (see
e.g.\ Figure 9 of Paper I), the effective sound speed is somewhat {\it lower},
perhaps surprisingly as the turbulent velocity dispersions of disk galaxies are
often thought to be {\it due\/} to feedback. We discuss this point further in
\autoref{feedback_conclusions}.

\begin{figure}
\plotone{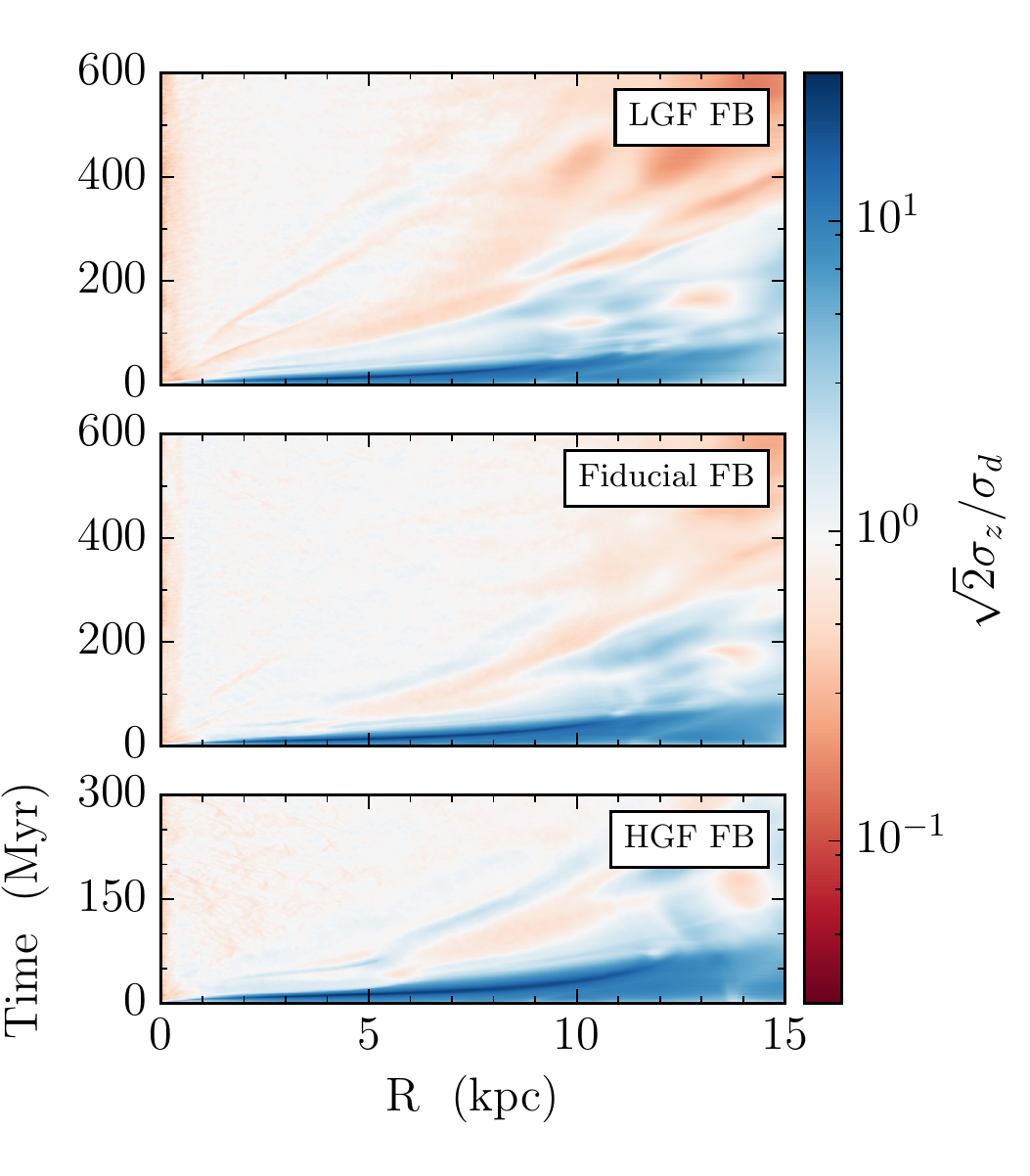}
\caption{The gas velocity dispersion anisotropy as a function of radius and
  time.}
\label{radius_time_gas_velocity_dispersion_ratio_feedback}
\end{figure}

On the other hand, a major effect of feedback is to make the turbulent velocity
field isotropic across the bulk of our simulated galaxies, with
$\sqrt{2}\sigma_{v,z}/\sigma_{v,d} \simeq 1$ at all radii. We can also see this
as a function of time in
\autoref{radius_time_gas_velocity_dispersion_ratio_feedback}.

Compared to the runs with no feedback, there is more power in the out-of-disk
component of the turbulent velocity dispersions, producing substantially larger
gas scale heights in these simulations.  Rather than all of the gas collapsing
into an extremely thin disk, gas extends both above and below the disk as it
gets thrown out of the midplane by feedback.  Typically, we find scale heights
of several hundred parsecs.

\subsubsection{Toomre Q}

\label{feedback_gravitational_instability}

\begin{figure}
\plotone{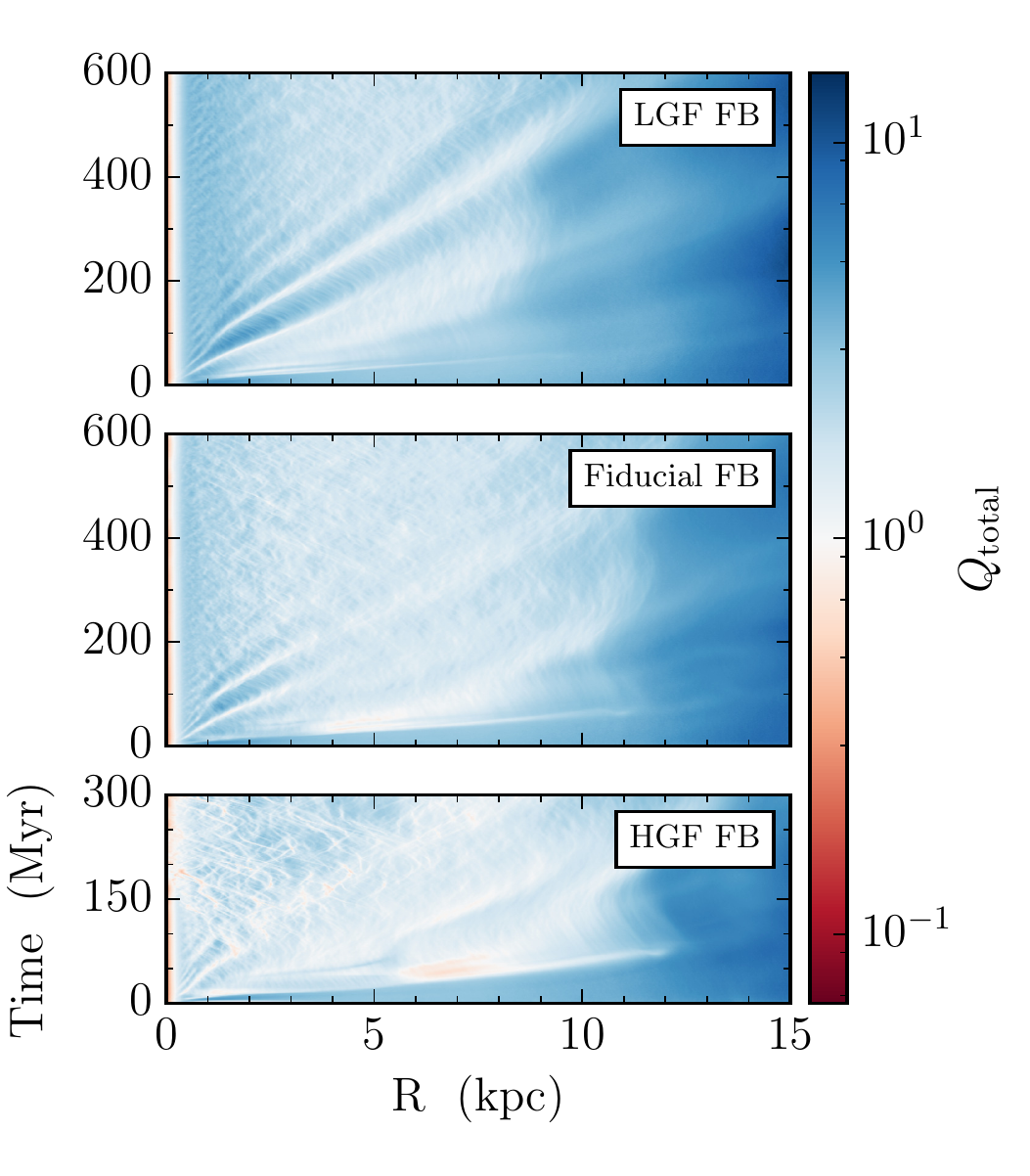}
\caption{The combined Toomre Q parameter as a function of radius and time.}
\label{radius_time_gas_total_toomre_q_feedback}
\end{figure}

\begin{figure}
\epsscale{1.1}
\plotone{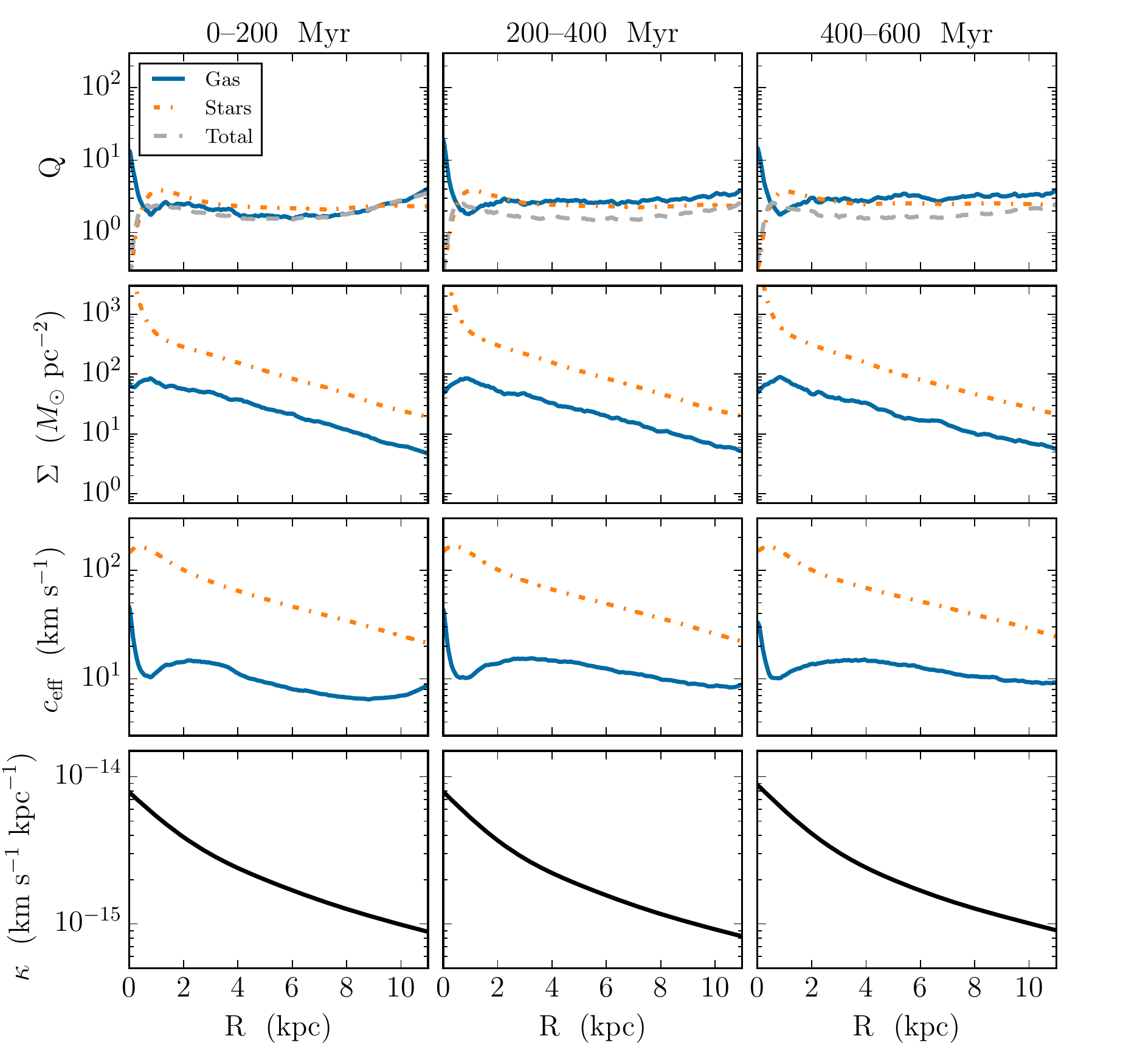}
\epsscale{1.0}
\caption{The gas, stellar, and total Toomre Q parameter (top row), gas and
  stellar surface density (second row), gas and stellar effective sound speed
  (third row) and epicyclic frequency (bottom row) for the low gas fraction
  simulation.}
\label{q_summary_plot_feedback}
\end{figure}

The susceptibility of a disk to gravitational instability can be characterized
via the Toomre $Q$ parameter. We are able to calculate a combined
$Q_{\rm total}$ that includes the separate contribution to the stability of the
disk from both gas and stars, and accounts for the finite thickness of the
gaseous and stellar disks \citep{romeo11}. We found in Paper I that, in the
absence of feedback, the galaxy initially stabilizes at
$Q_{\rm total} \approx 1$, but that as the simulation proceeds and gas is
exhausted, $Q_{\rm total}$ gradually rises.

We see from \autoref{radius_time_gas_total_toomre_q_feedback} that in our runs
including stellar feedback, $Q_{\rm total}$ is remarkably stable, showing little
variation with radius or time. There is some modulation at the factor-of-two
level due to surface density fluctuations --- transient rings early in the
simulation and spiral density waves later on --- but we do not see order of
magnitude variations as in the runs without feedback.

Stellar feedback is able to slow down the runaway gravitational instability that
would otherwise take hold by efficiently destroying gravitationally bound
clouds. The disk attains a quasi-stable state, with $Q_{\rm total} \gtrsim 1$
throughout the disk.  Both the gaseous and stellar components show little
variation over the course of the simulation, as we see in
\autoref{q_summary_plot_feedback}. The gaseous and stellar surface density
profile remain smooth, varying exponentially with radius at all times.

We see that stellar feedback is necessary to prevent the runaway fragmentation
of star forming gaseous disks, although the story is somewhat different from the
naive expectation that feedback drives turbulence.  Instead, feedback moderates
the consumption of gas, keeping $Q_{\rm total}$ close to unity, rather than
being driven to a high value by the exhaustion of gas. Crucially, and contrary
to assumptions commonly made in theoretical models, this means that feedback
actually lowers rather than increases $Q_{\rm total}$ on galaxy-average scales.
Only on the scales of individual molecular clouds, where simulations without
feeback reach $Q_{\rm total} \lesssim 1$, does feedback increase
$Q_{\rm total}$, by dispersing dense clouds via supernova blastwaves.

\subsubsection{Mass Transport}

Finally, we focus on the radial mass transport rate in the gaseous disk. In the
simulations without feedback, we found that mass transport was dominated by
N-body migration of individual giant clumps. We also found that there was a net
radial inflow of gas. While the inflow rate was sufficient to power the observed
star formation rates of Milky Way-like galaxies, it was insufficient to supply
the rapid star formation in the simulations without feedback.

\label{feedback_mass_transport}

\begin{figure}
\plotone{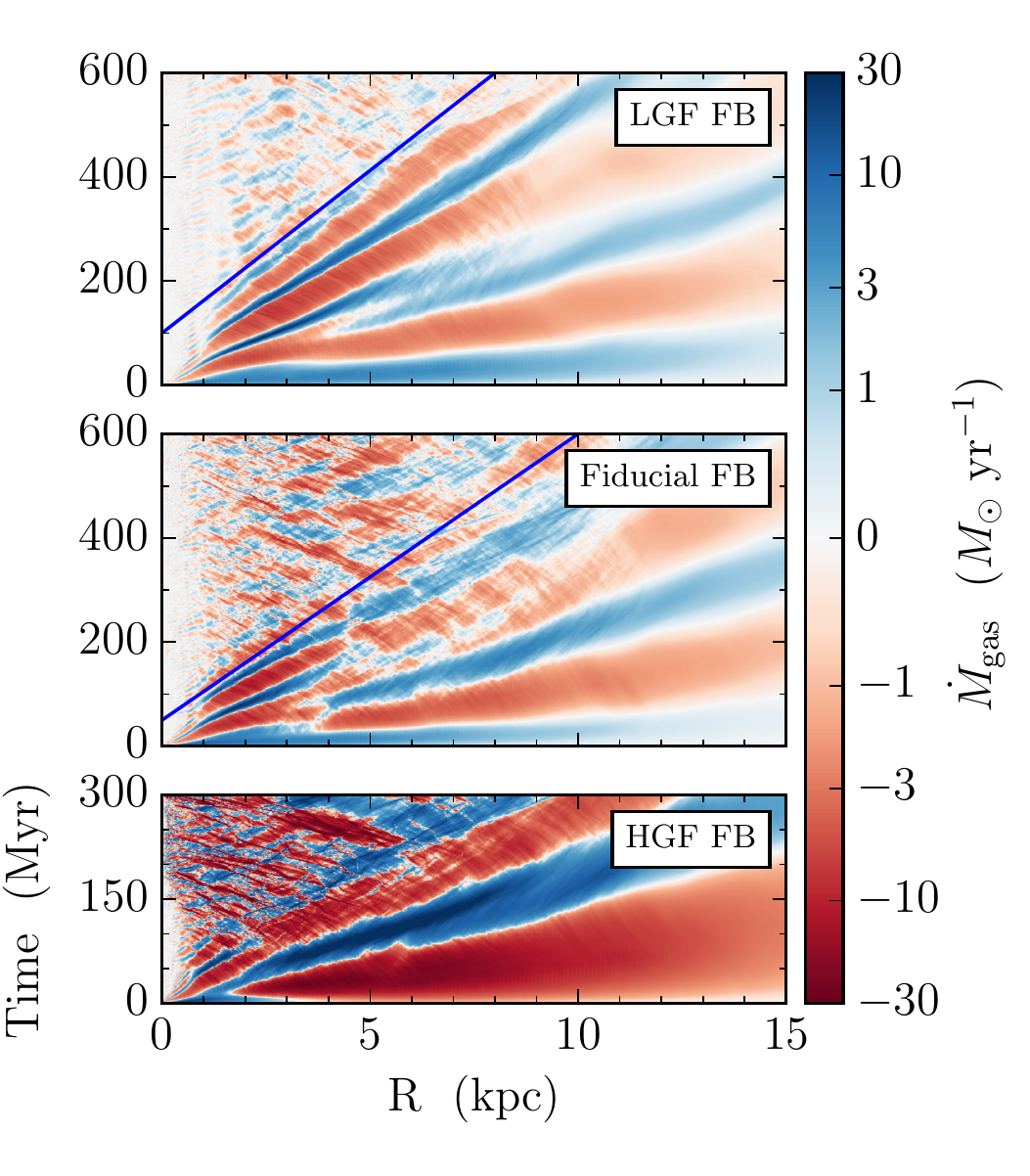}
\caption{The radial mass flux as a function of radius and time. Negative mass
  fluxes correspond to radial inward flow, while positive mass fluxes correspond
  to radial outward flow. The blue lines indicate the wedge averaging region
  used to produce \autoref{velocity_summary_plot_feedback} and
  \autoref{mass_flux_summary_feedback}.}
\label{radius_time_gas_mass_flux_feedback}
\end{figure}

The runs with star formation feedback show a very different history. In
\autoref{radius_time_gas_mass_flux_feedback}, we show the time evolution of the
radial mass flux.  Compared to the runs without feedback, the early evolution is
similar. At early times, the gas mass flux is dominated by rings of inward and
outward flux. Since we are mostly interested in the behavior of the disks after
they have settled down into statistical equilibrium, we do not consider the
portions of radius-time phase space that are strongly influenced by initial
transients (indicated by the blue lines in
\autoref{radius_time_gas_mass_flux_feedback}).  In the inner disk
($R \lesssim \unit[2]{kpc}$ for the low gas fraction case and
$R \lesssim \unit[1]{kpc}$ for the fiducial simulation), there is very little
mass transport. This is not surprising: our galaxies have realistic stellar
bulges, and as a result the shape of the rotation curve strongly suppresses
gravitational instability in the very inner portion of the disk. We found
similarly low rates of transport in this region for the no feedback simulations
in Paper I. At intermediate radii
$\unit[1.5]{kpc} \lesssim R \lesssim \unit[8]{kpc}$, the story is quite
different.  The mass flux shows alternating patterns of inward and outward
transport, which we associate with the formation of spiral arms. We cannot say
whether the high gas fraction case shows similar behavior, since it has not run
long enough.

\begin{figure}
\plotone{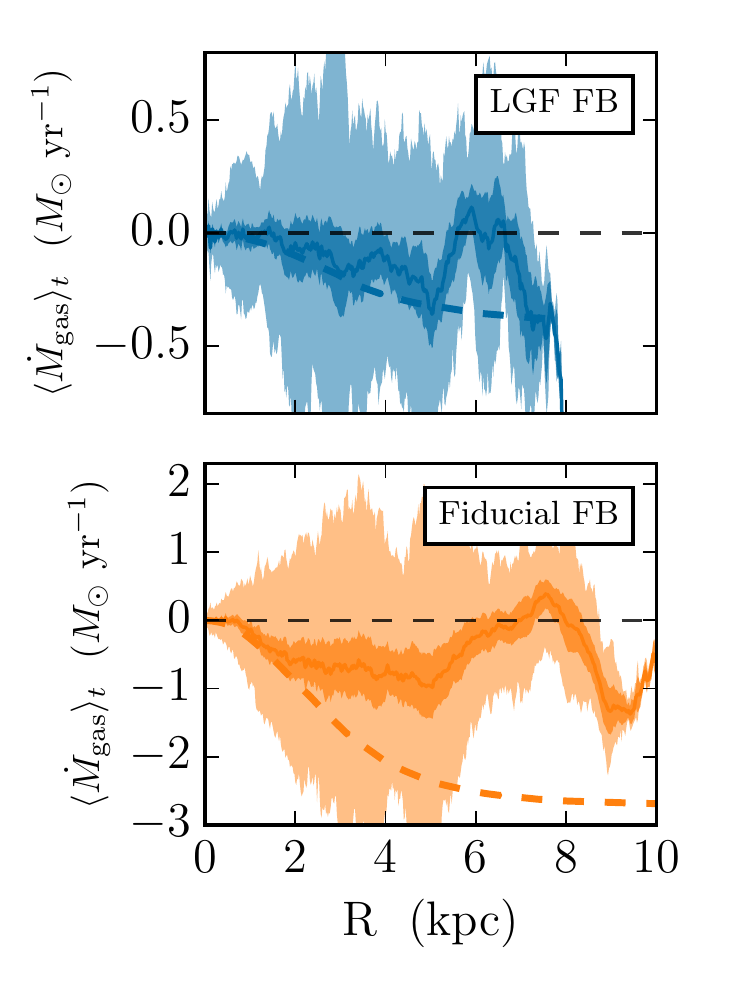}
\caption{The time-averaged radial mass flux as a function of radius for the low
  gas fraction (top panel) and fiducial (bottom panel) simulations. At a given
  radius, the light-shaded region encloses 67\% of the mass flux measurements
  over the course of the simulation. The dark-shaded region encloses the
  1-$\sigma$ confidence region for the mean mass flux, which we compute by
  taking the standard deviation of the mass flux measurements and dividing by
  the square root of the number of uncorrelated measurements. This measure of
  the uncertainty in the mean mass flux assumes a correlation timescale of
  \unit[10]{Myr}, which was measured by eye based on the autocorrelation of the
  mass flux measurements as a function of time at several radii. The dark, solid
  line is the mean mass flux. The large variations in mass flux visibly evident
  in \autoref{radius_time_gas_mass_flux_feedback} correspond to the significant
  vertical extent of the light-shaded region. The narrowness of the dark-shaded
  region indicates that inward net mass transport through the disk is
  statistically significant. We also plot the negative of the time-averaged
  radially accumulated star formation rate (dot-dashed lines) for comparison. At
  radii where the dashed line and the solid line lie on top of each other,
  radial mass transport is sufficient to fuel all star formation within that
  radius.\\}
\label{mass_flux_summary_feedback}
\end{figure}

The quantity of greatest interest from the standpoint of star formation fueling
is not the instantaneous transport rate, but its long term average.  We have
therefore calculated the time-averaged mass flux through each radial ring, which
we present in \autoref{mass_flux_summary_feedback}. The time averaging is
conducted in the wedge-shaped regions of radius-time parameter space above the
blue lines in \autoref{radius_time_gas_mass_flux_feedback}. This is done to
avoid including the time when the disk in undergoing initial collapse and
transient behavior. See Paper I for more details.

We see that the rates of inward mass transport, $\sim 0.1$ $M_\odot$ yr$^{-1}$
in the low gas fraction simulation and $\sim 1$ $M_\odot$ yr$^{-1}$ in the
fiducial one, are comparable to or perhaps slightly smaller what we found in the
simulations without feedback (c.f.~Figure 7 of Paper I). Thus we conclude that
the presence or absence of feedback only modestly affects the rate of mass
transport through a galactic disk.

For comparison, we also compute the cumulative time-averaged star formation rate
interior to each radius, defined simply as the total mass of stars formed whose
formation point is interior to that radius in the same wedge-averaging region,
divided by the time over which we measure star formation, which is a function of
radius. In \autoref{mass_flux_summary_feedback} we show the negative of this
quantity in order to ease comparison with the time-averaged mass flux.

At a fixed radius, if the radially cumulative star formation rate is equal to
the net inward mass flux, then the mass flux at that radius is sufficient to
supply all of the star formation within that radius.  For the low gas fraction
simulation, we see that over the bulk of the star forming disk
($R \lesssim \unit[5]{kpc}$), radial mass transport is sufficient to supply the
bulk of the star formation.

For the fiducial simulation, the mass transport rate is only sufficient to
supply the star formation for $R \lesssim \unit[2]{kpc}$. Between
$\unit[2]{kpc} \lesssim R \lesssim \unit[5]{kpc}$, the radial mass transport
rate slowly increases in magnitude from $\sim \unit[0.5]{\msun\ yr^{-1}}$ up to
$\sim \unit[1.0]{\msun\ yr^{-1}}$ while the radially cumulative star formation
rate increases from $\sim \unit[0.5]{\msun\ yr^{-1}}$ to
$\sim \unit[2.5]{\msun\ yr^{-1}}$. In this region the inward flow of gas is
insufficient to sustain star formation, and the gas supply would eventually be
substantially depleted if we followed the simulation beyond
\unit[600]{Myr}. Further out, the radially cumulative star formation rate only
slowly increases to $\sim \unit[2.8]{\msun\ yr^{-1}}$, while the radial mass
transport rate approaches zero, and even becomes slightly positive near
$R = \unit[8]{kpc}$. We caution that the mass transport rates at these radii
were measured using a small fraction of the simulation data we have on hand due
to the shape of the averaging region (see
\autoref{radius_time_gas_mass_flux_feedback}). The disk has experienced
comparatively little time after initial transient period at these radii so
running our simulations for a few more galactic dynamical times may shift the
average mass flux.

While the measured mass transport rates in our simulations are not sufficient to
supply all of the gas needed for star formation at intermediate radii in the
fiducial simulation, it is sufficient to supply a substantial fraction
($\sim 1/3$) of the necessary gas, slowing gas consumption and increasing the
depletion time. If we continued the simulation long enough for the gas fraction
to decrease to 10\%, comparable to the low gas fraction run, then it seems
likely that star formation and gas transport would reach full equilibrium as
they do in the low gas fraction case.  This equilibrium would likely persist
until the outer disk was drained of gas.

\section{Discussion and Conclusions}

\label{discussion_conclusions}

In this paper we have presented three simulations of Milky Way-like disk
galaxies under the influence of gravitational instability and star formation
feedback. By comparing these results to those we obtained in the absence of
stellar feedback (Paper I), we are able to separate out the roles of
gravitational instability and stellar feedback in determining the properties of
galactic disks, and we are able to study how mass transport might fuel star
formation in disk centers.

\subsection{The Effects of Feedback}

\label{feedback_conclusions}

We find that feedback is primarily responsible for preventing the interstellar
medium from becoming dominated by gravitationally-bound clouds, and instead
partitioning it into comparable masses of WNM, CNM, and molecular gas with a
relatively smooth morphology when averaged over $\sim$kpc scales.  When we
include feedback in our models, we find that our simulated galaxies reach a
balance between these phases that is in good agreement with values observed in
nearby Milky Way-like disks.  Feedback is also responsible for isotropizing the
turbulence in galactic disks and thereby pumping up their scale heights compared
to what would be expected in its absence. Most importantly, feedback is required
to suppress the star formation rate and produce disks with molecular gas
depletion times $t_{\rm dep} \sim \unit[2]{Gyr}$, comparable to what is observed
in local disks.

On the other hand, feedback is not the main agent determining either the
velocity structure or gravitational stability of the interstellar
medium. Contrary to naive expectations, including feedback actually reduces the
total velocity dispersion and Toomre Q parameter. In the absence of feedback,
gravitational instability is fully capable of maintaining large velocity
dispersions and preventing $Q_{\rm total}$ from dipping below unity. Feedback
does prevent the formation of local regions where $Q_{\rm total} < 1$ --- in the
simulations with no feedback these regions correspond to the locations of
massive star forming clouds. Feedback disrupts star forming clouds before enough
mass is able to accumulate locally to drive $Q_{\rm total}$ to values below
unity. However, when we compute global averages, the effect of feedback is to
lower $Q_{\rm total}$ and not raise it.

As we saw in \autoref{feedback_velocity_structure}, the time-averaged
azimuthally-averaged effective sound speed is \textit{suppressed} somewhat compared to
the simulations with no feedback. For example, compare
\autoref{velocity_summary_plot_feedback} with Figure 9 in Paper I. This can be
explained by examining the time evolution of the effective sound speed in both
simulations, as plotted in \autoref{q_summary_plot_feedback} and Figure 5 of
Paper I. In the first \unit[200]{Myr} of the simulations with no feedback, the
azimuthally averaged effective sound speed is initially very similar, albeit
somewhat suppressed, compared to the simulations with feedback. Later on, as the
gas supply is exhausted, the effective sound speed increases. The increase in
the effective sound speed works in concert with the decrease in the gas surface
density, increasing $Q_{\rm gas}$ throughout the star forming portion of the
gaseous disk.

In the simulations with feedback, the gas consumption timescale is
substantially longer, and the gas surface density profile does not vary much
over the course of the simulation. The lower turbulent velocity dispersions in
the simulation with feedback are therefore a reflection of the prolonged period
of marginal gravitational stability. Eventually, if the simulations were allowed
to run for another \unit[10$^9$]{yr}, and no new gas was added by cosmological
accretion, we would expect the gas surface density to
decrease, and turbulent velocity dispersion to increase, matching the behavior
seen in the simulations with no feedback.

This finding undermines the central assumption made in analytic models that
attempt to derive a star formation rate by deducing a value required to maintain
$Q_{\rm total} \approx 1$ \citep[e.g.,][]{faucher-giguere13}. It appears that
the star formation rate and the maintenance of marginal gravitational stability
are physically decoupled phenomena. It is also problematic for models that
derive a star formation rate from vertical hydrostatic equilibrium without
requiring that $Q_{\rm total} \approx 1$ \citep[e.g.,][]{ostriker10,
  ostriker11}; these models derive the star formation rate from vertical force
balance, which is reasonable in light of our finding that feedback is the main
agent responsible for setting galaxies' vertical scale heights. On the other
hand, these models also posit that galaxies with high surface densities reach
$Q_{\rm total} \ll 1$, which we find that gravitational instability always
prevents.

\subsection{Galactic Winds}

\label{galactic_winds}

Our simulations do not produce substantial galactic winds. This may be
surprising given the prominent role winds are thought to play in galaxy
evolution \citep[][and references therein]{veilleux05, peeples14}. We see two
possible explanations for why our simulations do not produce substantial winds.

First, recent simulations by the FIRE collaboration \citep{muratov15} ---
cosmological zoom-in simulations with a feedback recipe capable of producing 
strong winds --- indicate that galaxies with masses comparable to the
Milky Way do not produce substantial winds at low redshift. Going from high redshift
to low redshift, their m12i halo transitions from a bursty mode of star
formation to a mode where stars form at a steady rate. The authors of this study
conclude that the fall-off in the rate of mass ejected by winds is due to the
deepening of the gravitational potential, less concentrated star formation, and
less bursty star formation. All three effects conspire to reduce the
effectiveness of gas acceleration out of the halo. In a post-hoc analytical
investigation of the lack of winds seen at late times in the FIRE simulations,
\citet{hayward15} attribute the fall-off in the outflow rate to the decline in
the gas fraction. They predict that the ratio of the outflow rate to the star
formation rate falls off exponentially at low gas fractions. This may easily account
for the lack of appreciable outflows in the LGF and fiducial simulations.

To explain why we do not launch winds in the HGF simulation, and possibly also
for the others, our simulations may not have sufficient resolution to adequately
model the details of the wind acceleration process. In the theoretical models of
\citet{maclow88}, the expanding bubble of hot gas that drives the wind must last
long enough so that radiative losses cannot drain enough energy for the wind to
stall. In particular, the cooling timescale must be longer than the time for the
bubble radius to exceed a galactic scale height. In practice, this requires hot
gas, with temperatures exceeding \unit[$10^7$]{K}. The supernova bubbles in our
simulations do not create gas that is so hot. This is proximately due to the
implementation of our feedback recipe, which uses most of the energy of the
supernova explosions for direct momentum injection rather than the injection of
thermal energy. As discussed in section \autoref{sn_feedback}, we take this
approach to avoid the overcooling problem, namely that a single supernova
explosion must spread its energy over a volume of \unit[8000]{pc$^3$} given our
resolution, which will typically not heat the gas to a high enough temperature
to prevent it from cooling quickly. At the resolution of our simulations, it is
therefore difficult to avoid the result that our SNe produce relatively little
hot gas.

This issue is more prominent for us than it is for the FIRE simulations
\citep{muratov15} discussed above because, although their resolution is no
better than ours, they adopt a star
formation recipe that converts all gravitationally-bound gas into stars in a
single dynamical time. In contrast, we adopt an observationally-motivated recipe
whereby only $\sim 1\%$ of the mass per dynamical time is converted to stars
(see Paper 1 for details). This is relevant for galactic winds because it means
that the FIRE simulations have much lower densities in the regions where SNe
explode than we find, which in turn greatly reduces the resolution required to
properly capture the adiabatic phase of supernova remnant evolution.

\subsection{Implications for Star Formation Fueling}

Perhaps the most striking conclusion that can be drawn from our simulations from
the standpoint of galaxy evolution concerns star formation fueling. We find that
gravitational instability drives gas inflows through galactic disks regardless
of whether we include feedback in our simulations or not; feedback only very
modestly reduces the inflow rates, even as it completely disperses the large,
bound clumps that dominate our disks in the simulations without feedback. This
means that inflow is not limited to clumpy high-redshift disks
\citep[e.g.,][]{dekel09b}, but is instead a ubiquitous phenomenon that persists
to $z=0$. Once we include feedback in our simulations, thereby lowering the star
formation rates we measure, this inflow is sufficient to power all of the
present-day star formation in a Milky Way-mass galaxy with a 10\% gas fraction,
and to power all of the star formation in the core of a galaxy with a 20\% gas
fraction and a significant fraction of the star formation at larger radii. This
answers the question of why galaxy centers in Milky Way mass galaxies are
usually not devoid of gas and star formation. Even though these regions have
depletion times much less than a Hubble time, they can be re-supplied from
larger radii. This lowers the rate at which the gas fractions drop, and produces
a full equilibrium between consumption and infall once the gas fraction is low
enough, $\sim 10\%$. This equilibrium can presumably last as long as there is
sufficient gas available at large galactocentric radii, a condition that can be
satisfied for $\sim 1$ Hubble time or longer even in the absence of resupply
from outside the galaxy.

\acknowledgments\

This work utilized the Hyades supercomputer at the University of California
Santa Cruz, which is supported by the National Science Foundation through award
AST-1229745, and the Pleiades supercomputer, which is supported by the NASA
Advanced Supercomputing Division. The computations and analysis described in
this paper rely heavily on open source software packages, including
\texttt{Enzo}, \texttt{Python}, \texttt{yt}, \texttt{IPython}, \texttt{NumPy},
\texttt{SciPy}, \texttt{matplotlib}, \texttt{Cython}, \texttt{hdf5},
\texttt{h5py}, \texttt{scikit-image}, and \texttt{numexpr}.  We thank the
developer communities of these packages for their contributions. We also thank
the anonymous referee for providing comments that substantially improved the
quality of this manuscript. This work was supported by NSF graduate fellowships
(NJG and JCF), by NSF grants AST-0955300, AST-1405962 (MRK, NJG, and JCF), and
ACI-1535651 (NJG), by NASA TCAN grant NNX14AB52G (MRK, NJG, and JCF), by Hubble
Archival Research grant HST-AR-13909 (JCF and MRK), by grant DP16010100695 from
the Australian Research Council (MRK), and by the Gordon and Betty Moore
Foundation's Data-Driven Discovery Initiative through Grant GBMF4651 to Matthew
Turk.

\bibliography{./refs.bib}

\begin{thebibliography}{}
\expandafter\ifx\csname natexlab\endcsname\relax\def\natexlab#1{#1}\fi

\bibitem[{{Agertz} {et~al.}(2009){Agertz}, {Lake}, {Teyssier}, {Moore},
  {Mayer}, \& {Romeo}}]{agertz09}
{Agertz}, O., {Lake}, G., {Teyssier}, R., {et~al.} 2009, \mnras, 392, 294

\bibitem[{{Agertz} {et~al.}(2015){Agertz}, {Romeo}, \& {Grisdale}}]{agertz15}
{Agertz}, O., {Romeo}, A.~B., \& {Grisdale}, K. 2015, \mnras, 449, 2156

\bibitem[{{Bigiel} \& {Blitz}(2012)}]{bigiel12}
{Bigiel}, F., \& {Blitz}, L. 2012, \apj, 756, 183

\bibitem[{{Bigiel} {et~al.}(2008){Bigiel}, {Leroy}, {Walter}, {Brinks}, {de
  Blok}, {Madore}, \& {Thornley}}]{bigiel08}
{Bigiel}, F., {Leroy}, A., {Walter}, F., {et~al.} 2008, \aj, 136, 2846

\bibitem[{{Bigiel} {et~al.}(2011){Bigiel}, {Leroy}, {Walter}, {Brinks}, {de
  Blok}, {Kramer}, {Rix}, {Schruba}, {Schuster}, {Usero}, \&
  {Wiesemeyer}}]{bigiel11}
{Bigiel}, F., {Leroy}, A.~K., {Walter}, F., {et~al.} 2011, \apjl, 730, L13

\bibitem[{{Bonnell} {et~al.}(2013){Bonnell}, {Dobbs}, \& {Smith}}]{bonnell13}
{Bonnell}, I.~A., {Dobbs}, C.~L., \& {Smith}, R.~J. 2013, \mnras, 430, 1790

\bibitem[{{Bournaud} {et~al.}(2010){Bournaud}, {Elmegreen}, {Teyssier},
  {Block}, \& {Puerari}}]{bournaud10}
{Bournaud}, F., {Elmegreen}, B.~G., {Teyssier}, R., {Block}, D.~L., \&
  {Puerari}, I. 2010, \mnras, 409, 1088

\bibitem[{{Bresolin} {et~al.}(2012){Bresolin}, {Kennicutt}, \&
  {Ryan-Weber}}]{bresolin12}
{Bresolin}, F., {Kennicutt}, R.~C., \& {Ryan-Weber}, E. 2012, \apj, 750, 122

\bibitem[{{Bresolin} {et~al.}(2009){Bresolin}, {Ryan-Weber}, {Kennicutt}, \&
  {Goddard}}]{bresolin09}
{Bresolin}, F., {Ryan-Weber}, E., {Kennicutt}, R.~C., \& {Goddard}, Q. 2009,
  \apj, 695, 580

\bibitem[{{Cacciato} {et~al.}(2012){Cacciato}, {Dekel}, \&
  {Genel}}]{cacciato12}
{Cacciato}, M., {Dekel}, A., \& {Genel}, S. 2012, \mnras, 421, 818

\bibitem[{{Cen} \& {Ostriker}(1992)}]{cen92}
{Cen}, R., \& {Ostriker}, J.~P. 1992, \apjl, 399, L113

\bibitem[{{Cioffi} {et~al.}(1988){Cioffi}, {McKee}, \&
  {Bertschinger}}]{cioffi88}
{Cioffi}, D.~F., {McKee}, C.~F., \& {Bertschinger}, E. 1988, \apj, 334, 252

\bibitem[{{Daddi} {et~al.}(2010){Daddi}, {Elbaz}, {Walter}, {Bournaud},
  {Salmi}, {Carilli}, {Dannerbauer}, {Dickinson}, {Monaco}, \&
  {Riechers}}]{daddi10}
{Daddi}, E., {Elbaz}, D., {Walter}, F., {et~al.} 2010, \apjl, 714, L118

\bibitem[{{Dekel} {et~al.}(2009){Dekel}, {Sari}, \& {Ceverino}}]{dekel09b}
{Dekel}, A., {Sari}, R., \& {Ceverino}, D. 2009, \apj, 703, 785

\bibitem[{{Dobbs} {et~al.}(2011){Dobbs}, {Burkert}, \& {Pringle}}]{dobbs11b}
{Dobbs}, C.~L., {Burkert}, A., \& {Pringle}, J.~E. 2011, \mnras, 417, 1318

\bibitem[{{Dutton}(2012)}]{dutton12}
{Dutton}, A.~A. 2012, \mnras, 424, 3123

\bibitem[{{Faucher-Gigu{\`e}re} {et~al.}(2013){Faucher-Gigu{\`e}re},
  {Quataert}, \& {Hopkins}}]{faucher-giguere13}
{Faucher-Gigu{\`e}re}, C.-A., {Quataert}, E., \& {Hopkins}, P.~F. 2013, \mnras,
  433, 1970

\bibitem[{{Ferguson} \& {Clarke}(2001)}]{ferguson01}
{Ferguson}, A.~M.~N., \& {Clarke}, C.~J. 2001, \mnras, 325, 781

\bibitem[{{Field} {et~al.}(1969){Field}, {Goldsmith}, \& {Habing}}]{fgh69}
{Field}, G.~B., {Goldsmith}, D.~W., \& {Habing}, H.~J. 1969, \apjl, 155, L149

\bibitem[{{Forbes} {et~al.}(2012){Forbes}, {Krumholz}, \& {Burkert}}]{forbes12}
{Forbes}, J., {Krumholz}, M., \& {Burkert}, A. 2012, \apj, 754, 48

\bibitem[{{Forbes} {et~al.}(2014){Forbes}, {Krumholz}, {Burkert}, \&
  {Dekel}}]{forbes14}
{Forbes}, J.~C., {Krumholz}, M.~R., {Burkert}, A., \& {Dekel}, A. 2014, \mnras,
  438, 1552

\bibitem[{{Fraternali} {et~al.}(2013){Fraternali}, {Marasco}, {Marinacci}, \&
  {Binney}}]{fraternali13}
{Fraternali}, F., {Marasco}, A., {Marinacci}, F., \& {Binney}, J. 2013, \apjl,
  764, L21

\bibitem[{{Goldbaum} {et~al.}(2015){Goldbaum}, {Krumholz}, \&
  {Forbes}}]{goldbaum15a}
{Goldbaum}, N.~J., {Krumholz}, M.~R., \& {Forbes}, J.~C. 2015, \apj, 814, 131

\bibitem[{{Hayward} \& {Hopkins}(2015)}]{hayward15}
{Hayward}, C.~C., \& {Hopkins}, P.~F. 2015, ArXiv e-prints, arXiv:1510.05650

\bibitem[{{Ianjamasimanana} {et~al.}(2012){Ianjamasimanana}, {de Blok},
  {Walter}, \& {Heald}}]{ianjamasimanana12}
{Ianjamasimanana}, R., {de Blok}, W.~J.~G., {Walter}, F., \& {Heald}, G.~H.
  2012, \aj, 144, 96

\bibitem[{{Ianjamasimanana} {et~al.}(2015){Ianjamasimanana}, {de Blok},
  {Walter}, {Heald}, {Caldu-Primo}, \& {Jarrett}}]{ianjamasimanana15}
{Ianjamasimanana}, R., {de Blok}, W.~J.~G., {Walter}, F., {et~al.} 2015, \aj,
  arXiv:1506.04156, in press, arXiv:1506.04156

\bibitem[{{Joung} \& {Mac Low}(2006)}]{joung06}
{Joung}, M.~K.~R., \& {Mac Low}, M.-M. 2006, \apj, 653, 1266

\bibitem[{{Joung} {et~al.}(2009){Joung}, {Mac Low}, \& {Bryan}}]{joung09}
{Joung}, M.~R., {Mac Low}, M.-M., \& {Bryan}, G.~L. 2009, \apj, 704, 137

\bibitem[{{Katz}(1992)}]{katz92}
{Katz}, N. 1992, \apj, 391, 502

\bibitem[{{Kennicutt} \& {Evans}(2012)}]{kennicutt12}
{Kennicutt}, R.~C., \& {Evans}, N.~J. 2012, \araa, 50, 531

\bibitem[{{Kennicutt}(1998)}]{kennicutt98}
{Kennicutt}, Jr., R.~C. 1998, \apj, 498, 541

\bibitem[{{Kim} {et~al.}(2011{\natexlab{a}}){Kim}, {Kim}, \&
  {Ostriker}}]{kim11}
{Kim}, C.-G., {Kim}, W.-T., \& {Ostriker}, E.~C. 2011{\natexlab{a}}, \apj, 743,
  25

\bibitem[{{Kim} \& {Ostriker}(2015)}]{kim15}
{Kim}, C.-G., \& {Ostriker}, E.~C. 2015, \apj, 802, 99

\bibitem[{{Kim} {et~al.}(2011{\natexlab{b}}){Kim}, {Wise}, {Alvarez}, \&
  {Abel}}]{kim11b}
{Kim}, J.-h., {Wise}, J.~H., {Alvarez}, M.~A., \& {Abel}, T.
  2011{\natexlab{b}}, \apj, 738, 54

\bibitem[{{Kimm} \& {Cen}(2014)}]{kimm14}
{Kimm}, T., \& {Cen}, R. 2014, \apj, 788, 121

\bibitem[{{Kimm} {et~al.}(2015){Kimm}, {Cen}, {Devriendt}, {Dubois}, \&
  {Slyz}}]{kimm15}
{Kimm}, T., {Cen}, R., {Devriendt}, J., {Dubois}, Y., \& {Slyz}, A. 2015,
  \mnras, 451, 2900

\bibitem[{{Krumholz} \& {Burkert}(2010)}]{krumholz10}
{Krumholz}, M., \& {Burkert}, A. 2010, \apj, 724, 895

\bibitem[{{Krumholz}(2014)}]{krumholz14}
{Krumholz}, M.~R. 2014, \physrep, 539, 49

\bibitem[{{Krumholz} \& {McKee}(2005)}]{krumholz05}
{Krumholz}, M.~R., \& {McKee}, C.~F. 2005, \apj, 630, 250

\bibitem[{{Krumholz} {et~al.}(2008){Krumholz}, {McKee}, \& {Tumlinson}}]{kmt08}
{Krumholz}, M.~R., {McKee}, C.~F., \& {Tumlinson}, J. 2008, \apj, 689, 865

\bibitem[{{Krumholz} {et~al.}(2009){Krumholz}, {McKee}, \& {Tumlinson}}]{kmt09}
---. 2009, \apj, 693, 216

\bibitem[{{Leitherer} {et~al.}(2014){Leitherer}, {Ekstr{\"o}m}, {Meynet},
  {Schaerer}, {Agienko}, \& {Levesque}}]{leitherer14}
{Leitherer}, C., {Ekstr{\"o}m}, S., {Meynet}, G., {et~al.} 2014, \apjs, 212, 14

\bibitem[{{Leitherer} {et~al.}(1999){Leitherer}, {Schaerer}, {Goldader},
  {Delgado}, {Robert}, {Kune}, {de Mello}, {Devost}, \&
  {Heckman}}]{leitherer99}
{Leitherer}, C., {Schaerer}, D., {Goldader}, J.~D., {et~al.} 1999, \apjs, 123,
  3

\bibitem[{{Leroy} {et~al.}(2008){Leroy}, {Walter}, {Brinks}, {Bigiel}, {de
  Blok}, {Madore}, \& {Thornley}}]{leroy08}
{Leroy}, A.~K., {Walter}, F., {Brinks}, E., {et~al.} 2008, \aj, 136, 2782

\bibitem[{{Leroy} {et~al.}(2013){Leroy}, {Walter}, {Sandstrom}, {Schruba},
  {Munoz-Mateos}, {Bigiel}, {Bolatto}, {Brinks}, {de Blok}, {Meidt}, {Rix},
  {Rosolowsky}, {Schinnerer}, {Schuster}, \& {Usero}}]{leroy13}
{Leroy}, A.~K., {Walter}, F., {Sandstrom}, K., {et~al.} 2013, \aj, 146, 19

\bibitem[{{Mac Low} \& {McCray}(1988)}]{maclow88}
{Mac Low}, M.-M., \& {McCray}, R. 1988, \apj, 324, 776

\bibitem[{{Martizzi} {et~al.}(2015){Martizzi}, {Faucher-Gigu{\`e}re}, \&
  {Quataert}}]{martizzi15}
{Martizzi}, D., {Faucher-Gigu{\`e}re}, C.-A., \& {Quataert}, E. 2015, \mnras,
  450, 504

\bibitem[{{McKee} \& {Krumholz}(2010)}]{mckee10}
{McKee}, C.~F., \& {Krumholz}, M.~R. 2010, \apj, 709, 308

\bibitem[{{Meidt} {et~al.}(2013){Meidt}, {Schinnerer}, {Garc{\'{\i}}a-Burillo},
  {Hughes}, {Colombo}, {Pety}, {Dobbs}, {Schuster}, {Kramer}, {Leroy}, {Dumas},
  \& {Thompson}}]{meidt13a}
{Meidt}, S.~E., {Schinnerer}, E., {Garc{\'{\i}}a-Burillo}, S., {et~al.} 2013,
  \apj, 779, 45

\bibitem[{{Mo} {et~al.}(1998){Mo}, {Mao}, \& {White}}]{mo98}
{Mo}, H.~J., {Mao}, S., \& {White}, S.~D.~M. 1998, \mnras, 295, 319

\bibitem[{{Muratov} {et~al.}(2015){Muratov}, {Kere{\v s}},
  {Faucher-Gigu{\`e}re}, {Hopkins}, {Quataert}, \& {Murray}}]{muratov15}
{Muratov}, A.~L., {Kere{\v s}}, D., {Faucher-Gigu{\`e}re}, C.-A., {et~al.}
  2015, \mnras, 454, 2691

\bibitem[{{Olivier} {et~al.}(1991){Olivier}, {Primack}, \&
  {Blumenthal}}]{olivier91}
{Olivier}, S.~S., {Primack}, J.~R., \& {Blumenthal}, G.~R. 1991, \mnras, 252,
  102

\bibitem[{{Onodera} {et~al.}(2010){Onodera}, {Kuno}, {Tosaki}, {Kohno},
  {Nakanishi}, {Sawada}, {Muraoka}, {Komugi}, {Miura}, {Kaneko}, {Hirota}, \&
  {Kawabe}}]{ono10}
{Onodera}, S., {Kuno}, N., {Tosaki}, T., {et~al.} 2010, \apjl, 722, L127

\bibitem[{{Oppenheimer} \& {Dav{\'e}}(2008)}]{oppenheimer08}
{Oppenheimer}, B.~D., \& {Dav{\'e}}, R. 2008, \mnras, 387, 577

\bibitem[{{Ostriker} {et~al.}(2010){Ostriker}, {McKee}, \&
  {Leroy}}]{ostriker10}
{Ostriker}, E.~C., {McKee}, C.~F., \& {Leroy}, A.~K. 2010, \apj, 721, 975

\bibitem[{{Ostriker} \& {Shetty}(2011)}]{ostriker11}
{Ostriker}, E.~C., \& {Shetty}, R. 2011, \apj, 731, 41

\bibitem[{{Parravano} {et~al.}(2003){Parravano}, {Hollenbach}, \&
  {McKee}}]{par03}
{Parravano}, A., {Hollenbach}, D.~J., \& {McKee}, C.~F. 2003, \apj, 584, 797

\bibitem[{{Peeples} {et~al.}(2014){Peeples}, {Werk}, {Tumlinson},
  {Oppenheimer}, {Prochaska}, {Katz}, \& {Weinberg}}]{peeples14}
{Peeples}, M.~S., {Werk}, J.~K., {Tumlinson}, J., {et~al.} 2014, \apj, 786, 54

\bibitem[{{Petit} {et~al.}(2015){Petit}, {Krumholz}, {Goldbaum}, \&
  {Forbes}}]{petit15}
{Petit}, A.~C., {Krumholz}, M.~R., {Goldbaum}, N.~J., \& {Forbes}, J.~C. 2015,
  \mnras, 449, 2588, submitted, arXiv:1411.7585

\bibitem[{{Renaud} {et~al.}(2013){Renaud}, {Bournaud}, {Emsellem}, {Elmegreen},
  {Teyssier}, {Alves}, {Chapon}, {Combes}, {Dekel}, {Gabor}, {Hennebelle}, \&
  {Kraljic}}]{renaud13}
{Renaud}, F., {Bournaud}, F., {Emsellem}, E., {et~al.} 2013, \mnras, 436, 1836

\bibitem[{{Romeo} \& {Wiegert}(2011)}]{romeo11}
{Romeo}, A.~B., \& {Wiegert}, J. 2011, \mnras, 416, 1191

\bibitem[{{Scannapieco} {et~al.}(2006){Scannapieco}, {Tissera}, {White}, \&
  {Springel}}]{scannapieco06}
{Scannapieco}, C., {Tissera}, P.~B., {White}, S.~D.~M., \& {Springel}, V. 2006,
  \mnras, 371, 1125

\bibitem[{{Schruba} {et~al.}(2010){Schruba}, {Leroy}, {Walter}, {Sandstrom}, \&
  {Rosolowsky}}]{sch10}
{Schruba}, A., {Leroy}, A.~K., {Walter}, F., {Sandstrom}, K., \& {Rosolowsky},
  E. 2010, \apj, 722, 1699

\bibitem[{{Simpson} {et~al.}(2014){Simpson}, {Bryan}, {Hummels}, \&
  {Ostriker}}]{simpson14}
{Simpson}, C.~M., {Bryan}, G.~L., {Hummels}, C., \& {Ostriker}, J.~P. 2014,
  ArXiv e-prints, arXiv:1410.3822

\bibitem[{{Smith} {et~al.}(2011){Smith}, {Hallman}, {Shull}, \&
  {O'Shea}}]{smith11}
{Smith}, B.~D., {Hallman}, E.~J., {Shull}, J.~M., \& {O'Shea}, B.~W. 2011,
  \apj, 731, 6

\bibitem[{{Springel} {et~al.}(2005){Springel}, {Di Matteo}, \&
  {Hernquist}}]{springel05}
{Springel}, V., {Di Matteo}, T., \& {Hernquist}, L. 2005, \mnras, 361, 776

\bibitem[{{Springel} \& {Hernquist}(2003)}]{springel03}
{Springel}, V., \& {Hernquist}, L. 2003, \mnras, 339, 289

\bibitem[{{Stinson} {et~al.}(2006){Stinson}, {Seth}, {Katz}, {Wadsley},
  {Governato}, \& {Quinn}}]{stinson06}
{Stinson}, G., {Seth}, A., {Katz}, N., {et~al.} 2006, \mnras, 373, 1074

\bibitem[{{Tamburro} {et~al.}(2009){Tamburro}, {Rix}, {Leroy}, {Mac Low},
  {Walter}, {Kennicutt}, {Brinks}, \& {de Blok}}]{tamburro09}
{Tamburro}, D., {Rix}, H.-W., {Leroy}, A.~K., {et~al.} 2009, \aj, 137, 4424

\bibitem[{{Tasker} \& {Bryan}(2008)}]{tasker08}
{Tasker}, E.~J., \& {Bryan}, G.~L. 2008, \apj, 673, 810

\bibitem[{{The Enzo Collaboration} {et~al.}(2014){The Enzo Collaboration},
  {Norman}, {O'Shea}, {Abel}, {Wise}, {Turk}, {Reynolds}, {Collins}, {Wang},
  {Skillman}, {Smith}, {Harkness}, {Bordner}, {Kim}, {Kuhlen}, {Xu},
  {Goldbaum}, {Hummels}, {Kritsuk}, {Tasker}, {Skory}, {Simpson}, {Hahn},
  {Oishi}, {So}, {Zhao}, {Cen}, {Li}, \& {Enzo Collaboration}}]{enzo14}
{The Enzo Collaboration}, {Bryan}, G.~L., {Norman}, M.~L., {O'Shea}, B.~W.,
  {et~al.} 2014, \apjs, 211, 19

\bibitem[{{Thompson} {et~al.}(2005){Thompson}, {Quataert}, \&
  {Murray}}]{thompson05}
{Thompson}, T.~A., {Quataert}, E., \& {Murray}, N. 2005, \apj, 630, 167

\bibitem[{{Toomre}(1964)}]{toomre64}
{Toomre}, A. 1964, \apj, 139, 1217

\bibitem[{{V{\'a}zquez} \& {Leitherer}(2005)}]{vazquez05}
{V{\'a}zquez}, G.~A., \& {Leitherer}, C. 2005, \apj, 621, 695

\bibitem[{{Veilleux} {et~al.}(2005){Veilleux}, {Cecil}, \&
  {Bland-Hawthorn}}]{veilleux05}
{Veilleux}, S., {Cecil}, G., \& {Bland-Hawthorn}, J. 2005, \araa, 43, 769

\bibitem[{{Werk} {et~al.}(2011){Werk}, {Putman}, {Meurer}, \&
  {Santiago-Figueroa}}]{werk11}
{Werk}, J.~K., {Putman}, M.~E., {Meurer}, G.~R., \& {Santiago-Figueroa}, N.
  2011, \apj, 735, 71

\bibitem[{{Wolfire} {et~al.}(2003){Wolfire}, {McKee}, {Hollenbach}, \&
  {Tielens}}]{wolfire03}
{Wolfire}, M.~G., {McKee}, C.~F., {Hollenbach}, D., \& {Tielens}, A.~G.~G.~M.
  2003, \apj, 587, 278

\bibitem[{{Yang} \& {Krumholz}(2012)}]{yang12}
{Yang}, C.-C., \& {Krumholz}, M. 2012, \apj, 758, 48

\end{thebibliography}

\end{document}